\RequirePackage{lineno}
\documentclass[aps,twocolumn,floatfix,showpacs,superscriptaddress,groupedaddress]{revtex4-1}  
\usepackage{graphicx}  
\usepackage{subfigure} 
\usepackage{bm}        
\usepackage{amsmath}   
\usepackage{cancel} 
\usepackage{verbatim} 
\usepackage[percent]{overpic}

\begin{document}

\title{Enhanced Dynamical Stability with Harmonic Slip-stacking}
\author{Jeffrey Eldred}
\affiliation{Fermi National Accelerator Laboratory, Batavia, Illinois 60510, USA}
\author{Robert Zwaska}
\affiliation{Fermi National Accelerator Laboratory, Batavia, Illinois 60510, USA}
\date{\today}

\begin{abstract}

We develop a configuration of radio-frequency (rf) cavities to dramatically improve the performance of slip-stacking. Slip-stacking is an accumulation technique used at Fermilab to nearly double proton intensity by maintaining two beams of different momenta in the same storage ring. The two particle beams are longitudinally focused in the Recycler by two 53 MHz 100 kV rf cavities with a small frequency difference between them. We propose an additional 106 MHz 20 kV rf cavity with a frequency at the double the average of the upper and lower main rf frequencies. We show the harmonic rf cavity cancels out the resonances generated between the two main rf cavities and we derive the relationship between the harmonic rf voltage and the main rf voltage. We find the area factors that can be used to calculate the available phase space area for any set of beam parameters without individual simulation. We establish Booster beam quality requirements to achieve 99\% slip-stacking efficiency. We measure the longitudinal distribution of the Booster beam and use it to generate a realistic beam model for slip-stacking simulation. We demonstrate that the harmonic rf cavity can not only reduce particle loss during slip-stacking, but also reduce the final longitudinal emittance.
\end{abstract}

\pacs{29.20.dk, 02.60.Cb, 29.27.-a}
\maketitle

\section*{Introduction}

Improving proton beam power is a research priority at Fermilab and slip-stacking will be a critical part of high-intensity operation at Fermilab for the foreseeable future\cite{Prebys,Adamson,DUNE}. The high-intensity 120 GeV proton beam is delivered from the Main Injector to a carbon target for Neutrinos at Main Injector (NuMI) experiments \cite{Minos,Minerva,Nova} and to a hydrogen-deuterium target for the SeaQuest experiment~\cite{SeaQuest}. The Fermilab Proton Improvement Plan-II (PIP-II)~\cite{PIP} entails a 70\% increase in beam power and also requires loss rates during slip-stacking to decrease proportionately in order to limit activation. Loss studies indicate that single-particle dynamics associated with slip-stacking are a dominant source of particle loss~\cite{Dis}. Recent work in the single-particle longitudinal dynamics of slip-stacking~\cite{Eldred} demonstrated that the slip-stacking loss rate could be significantly reduced by upgrading the Fermilab Booster cycle-rate from 15 Hz to 20 Hz. The 20 Hz Booster cycle-rate has subsequently been incorporated into the PIP-II proposal~\cite{Derwent}. 

In this paper, we demonstrate a new method of reducing particles loss during slip-stacking that could be achieved using a 106 MHz 20 kV harmonic rf cavity in the Fermilab Recycler. High-energy particle accelerators have achieved a variety of dynamical effects by using one or more rf cavities operating at a multiple of the main rf frequency. The application of harmonic rf cavity we propose is novel because it does not operate at precise multiple of any one rf cavity, but rather it operates at twice the average of the upper and lower frequency main rf cavities. We refer to this modification of slip-stacking as ``harmonic slip-stacking''. With harmonic slip-stacking, the upgrade of the Booster cycle-rate is no longer necessary for the 120-GeV program and the primary consideration would be the benefit to the 8-GeV program~\cite{BooNE,mu2e,g-2}.

In this paper, we adopt and extend the tools for analyzing the dynamical stability of slip-stacking that were introduced in \cite{Eldred}. Our analysis finds that harmonic slip-stacking can increase the stable phase-space area by 50\% relative to conventional slip-stacking. Further, a simulation based on a realistic model of the beam finds a reduction in losses by a factor of 20 and a reduction in the final longitudinal emittance by 5\%. These results indicate a performance that would far exceed the slip-stacking loss requirements of the Fermilab PIP-II upgrade.

Additionally, slip-stacking ion beams in the Super Proton Synchrotron (SPS) is part of the baseline scenario for the Large Hadron Collider (LHC) Injector Upgrade (LIU)~\cite{LIU,Argyropoulos}. Although there is currently no 400MHz rf cavity in the SPS~\cite{Valuch,Mastoridis}, the results of this paper imply that such a cavity would improve the efficiency of slip-stacking in the SPS and lower the final longitudinal emittance after slip-stacking.

\section*{Background}
Slip-stacking is a particle accelerator configuration that permits two high-energy particle beams of different momenta to use the same transverse space in a cyclic accelerator (see \cite{Dis,MacLachlan,Brown}). The two beams are longitudinally focused by two rf cavities with a small frequency difference between them. Each beam is synchronized to one rf cavity and perturbed by the other rf cavity.

\begin{figure}[htp]
	\centering
    \includegraphics[scale=0.24]{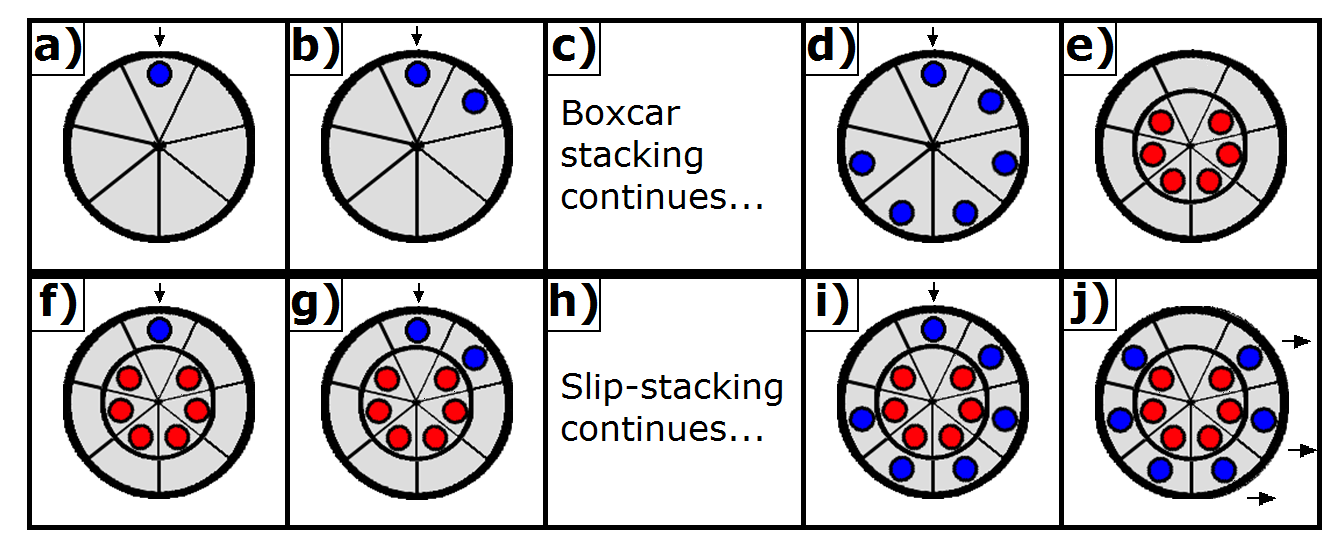}
  \caption{The Booster batch is represented by the circles and the Recycler (or Main Injector) is represented by the seven-sector wheel. {\bf a-d)} Six Booster batches injected each Booster cycle (boxcar stacking). {\bf e)} The rf frequency is gradually lowered in between the sixth and seventh batch injection. {\bf f-i)} Subsequent batches are injected into the gap left by the first six and gradually overlap. {\bf j)} When the first six and last six batches are aligned, the batches are extracted to the Main Injector (if needed) and both beams are accelerated as one.}
  \label{SmallM}
\end{figure}

For slip-stacking at Fermilab, the two azimuthal beam distributions are manipulated as a consequence of their difference in rf frequency. As shown in Fig.~\ref{SmallM}, the two beams are injected on separated portions of azimuth with a small frequency difference and overlap gradually. When the cyclic accelerator is filled and the azimuthal distribution of the two beams coincide, the two beams are accelerated together by an rf system operating at the average frequency. The potential beam intensity of a synchrotron is doubled through the application of this technique.

The rate at which the two slip-stacking beams slip with respect to each other must be synchronized with the rate that beam is injected. The difference between the two rf frequencies $\Delta f$ must be equal to the product of the harmonic number of the Booster rf $h_{B}$ and the cycle rate of the Booster $f_{B}$. For a 15-Hz Booster cycle-rate we have $\Delta f = h_{B} f_{B} = 1260$ Hz and for a 20-Hz Booster cycle-rate we have $\Delta f = h_{B} f_{B} = 1680$ Hz.

Prior work in the single-particle dynamics of slip-stacking can be found in \cite{Mills,Boussard,MacLachlan,Eldred}. Fermilab has implemented slip-stacking operationally since 2004 \cite{MacLachlan,SeiyaBC,Brown}. Initially, slip-stacking took place in the Fermilab Main Injector and now slip-stacking is operating in the Fermilab Recycler~\cite{Adamson}.

Beam-loading effects can impact the effectiveness of slip-stacking. A summary of beam-loading research can be found in \cite{Eldred} and draws upon work conducted for the slip-stacking in the Fermilab Main Injector \cite{SeiyaB,DeyKK,Dey,Madrak}. Slip-stacking influences the transverse dynamics of the Recycler by impacting the linear charge density and the chromatic tune-spread~\cite{Ainsworth,Balbekov}.

\section*{Slip-stacking Dynamics}

The longitudinal motion of a particle under the influence of an rf system is described by phase-space coordinates $\phi$ and $\delta$~\cite{SYBook}. The coordinate $\phi$ is the phase of the particle relative to the resonating electromagnetic field in the rf system. The coordinate $\delta$ is the fractional deviation from the reference momentum; $\delta=0$ corresponds to a particle whose revolution frequency $f_{rev}$ is a subharmonic of the frequency of the rf system $f_{rf} = h f_{rev}$. The phase-slip factor $\eta$ is used to describe how the revolution period $T$ changes with $\delta$ and is given by $\displaystyle \eta \delta = \Delta T / T$~\cite{SYBook}.

The equations of motion associated with the trajectory of a single particle under the influence of a stationary rf system \cite{SYBook} are given by
\begin{equation} \label{dif1srf}
\dot{\phi} = 2 \pi f_{rev} h \eta \delta,~\dot{\delta} = f_{rev} \frac{eV}{\beta^{2}E}\sin(\phi)
\end{equation}
$V$ is the effective voltage of the rf cavity, $e$ is the charge of the particle, $\beta = v/c$ is the velocity fraction of the speed of light, $E$ is the total energy of the particle.

The corresponding second-order equation of motion is
\begin{equation} \label{dif2srf}
\ddot{\phi} = - \omega_{s}^{2} \sin(\phi).
\end{equation}
where $\displaystyle \omega_{s} = 2\pi f_{rev} \sqrt{\frac{heV\eta}{2\pi\beta^{2}E}}$ is the synchrotron frequency (see \cite{SYBook}). The motion of the particle is longitudinally focused and $\omega_{s}$ is the frequency of small oscillations. At large amplitude, the oscillation frequency is given by $\omega_{s}(1+\sigma)$ where $\sigma$ is the amplitude-dependent tune-shift.

For slip-stacking, there is an rf phase associated with each of two cavities. Without loss of generality, we adopt the frame of reference of a particle synchronized with upper rf frequency and add the influence of the lower rf cavity. If the frequency difference between the two rf cavities is $\Delta f$ then the phase of the second cavity advances at $2 \pi \Delta f t$ relative to the phase of the first cavity. We define $\omega_{\phi} = 2 \pi \Delta f$, the phase-slipping frequency. The equations of motions for a single particle under the influence of two main rf cavities are given by
\begin{align} \nonumber 
\dot{\phi} &= 2 \pi f_{rev} h \eta \delta \\ \label{dif1ss}
\dot{\delta} &= f_{rev} \frac{eV}{\beta^{2}E} \big[\sin(\phi) + \sin(\phi - \omega_{\phi} t )\big].
\end{align}
The corresponding second-order equation of motion is expanded
\begin{align} \nonumber
\ddot{\phi} = - \omega_{s}^{2} \big[\sin(\phi) & + \sin(\phi) \cos(\omega_{\phi} t ) \\ \label{dif2ss}
& - \cos(\phi) \sin(\omega_{\phi} t )\big].
\end{align}

In the rapid slipping limit $\omega_{\phi} \gg \omega_{s}$ the perturbation from the second rf cavity averages out rapidly and Eq.~\ref{dif2ss} approaches Eq.~\ref{dif2srf}. Similarly, from the frame of reference of particles synchronized to the second rf cavity, the perturbation from the first rf cavity averages out rapidly. At more moderate values of $\omega_{\phi}$ relative to $\omega_{s}$, the perturbation effect complicates the motion and reduces the stable phase-space area. This effect is quantified by the slip-stacking parameter $\alpha_{s}$~\cite{Dis,Eldred,MacLachlan,Mills,Boussard} defined:
\begin{equation} \label{as} 
\alpha_{s} =\frac{\omega_{\phi}}{\omega_{s}}.
\end{equation}

A perturbation analysis of the trajectory of slip-stacking particles~\cite{Eldred} shows that the influence of the second cavity creates a series of parametric resonances. The synchrotron tune-shift $\sigma$ depends both on the synchrotron amplitude $\rho$ and the slip-stacking parameter perturbation $\alpha_{s}^{-2}$. When the synchrotron oscillation frequency (with the tune-shift) is a rational multiple of the phase-slipping frequency $m \omega_{s}(1+\sigma) = n \omega_{\phi}$ an uncontrolled growth term appears: $\ddot{\phi} \propto \rho^{m} \alpha_{s}^{-2(n-1)}$.

Equation~\ref{dif1srf} is equivalent to the simple pendulum and Eq.~\ref{dif1ss} is equivalent to the driven pendulum~\cite{Eldred}, but there is no clear pendulum analogue for harmonic slip-stacking.

\section*{Slip-stacking Dynamics with Harmonic rf}

For conventional slip-stacking in the Recycler, the two beams are maintained by two 53 MHz rf cavities, the upper rf frequency at $f_{0} +\Delta f /2$ and the lower rf frequency at $f_{0} - \Delta f /2$. For harmonic slip-stacking in the Recycler, a 106 MHz rf cavity would operate at twice the average frequency $2 f_{0}$. The harmonic rf cavity is not synchronized to either of the two beam, but helps keep both beams synchronized with their corresponding main rf cavity. We add a new term to Eq.~\ref{dif1ss} corresponding to this new rf cavity
\begin{alignat}{2} \nonumber 
\dot{\phi} &= 2 \pi f_{rev} h \eta \delta && \\ \nonumber
\dot{\delta} &= f_{rev} \frac{eV_{M}}{\beta^{2}E} \big[\sin(\phi) &&+ \sin(\phi - \omega_{\phi} t ) \\ \label{dif1hss}
& &&+ \lambda \sin(2\phi - \omega_{\phi} t )\big].
\end{alignat}
$V_{M}$ is the main rf voltage and $\lambda$ is the ratio between the harmonic rf voltage and main rf voltage $\lambda = V_{H} / V_{M}$. For a negative value of $\lambda$, the harmonic rf cavity defocuses at the phase that is the average of the focusing phases for the upper and lower rf cavity. Consequently the harmonic rf cavity partially counteracts the perturbation effect that the upper and lower rf cavity have on each other. In this section, we demonstrate that a negative value of $\lambda$ reduces slip-stacking resonance terms.

We write the second-order equation of motion corresponding to Eq.~\ref{dif1hss} and use a Taylor series to expand $\sin\phi$ and $\cos\phi$ as polynomials:
\begin{align} \nonumber
\ddot{\phi} = - \omega_{s}^{2} \Bigg[ & \sum_{k=0}^{\infty} \frac{(-1)^{k}}{(2k+1)!} \phi^{2k+1} \\ \nonumber
& + \sum_{k=0}^{\infty} \frac{(-1)^{k}}{(2k+1)!} \phi^{2k+1} \left(1 + \lambda 2^{2k+1}\right) \cos(\omega_{\phi} t) \\ \label{dif2hss}
& - \sum_{k=0}^{\infty} \frac{(-1)^{k}}{(2k)!} \phi^{2k} \left(1 + \lambda 2^{2k}\right) \sin(\omega_{\phi} t) \Bigg].
\end{align}
To understand the role of $\lambda$ in Eq.~\ref{dif2hss}, we consider the value of ${\phi^{m} \left[1 + \lambda 2^{m}\right]}$ in the case where ${\lambda = - 2^{-p}}$. For ${p=m}$ the coefficient is completely canceled, for ${p > m}$ the bracketed term is positive and less than 1, and for ${p < m}$ the bracketed term is negative. The $\phi^{m}$ term generates the lowest order contribution to the ${m \omega_{s}(1+\sigma) = \omega_{\phi}}$ resonance. Numerical studies indicate that a negative value of $\lambda$ partially counteracts the slip-stacking perturbation on the synchrotron tune shift $\sigma$.

This suggests a natural scaling of $\lambda$ with $\alpha_{s}$. Suppose that for some value of $\alpha_{s}$ there is some optimal value of $\lambda$ for which $\lambda = - 2^{-m}$ cancels some appropriate resonance ${m = \omega_{\phi} /[\omega_{s}(1+\sigma)]}= \alpha_{s}/(1+\sigma) $. We approximate $\sigma$ as a constant $\sigma_{0}$ and reduce $\lambda$ to a function of a single variable ${\lambda = - 2^{-\alpha_{s}/(1+\sigma_{0})}}$. We rewrite this expression for $\lambda$ in the simple exponential form
\begin{align} \label{fitform}
\lambda = - e^{-\xi \alpha_{s}}
\end{align}
with $\xi = \ln(2)/(1+\sigma_{0})$ a constant to be determined empirically. We expect this functional dependence to be valid at high values of $\alpha_{s}$ where the slip-stacking perturbation on $\sigma$ is weak and the slip-stacking resonances are close. In the next section, we show that the value of $\lambda$ which maximizes the phase-space area follows the form given in Eq.~\ref{fitform} for high values of $\alpha_{s}$.

\section*{Stable Phase-space Area}
\begin{figure}[htp]
	\centering
    \includegraphics[scale=0.4]{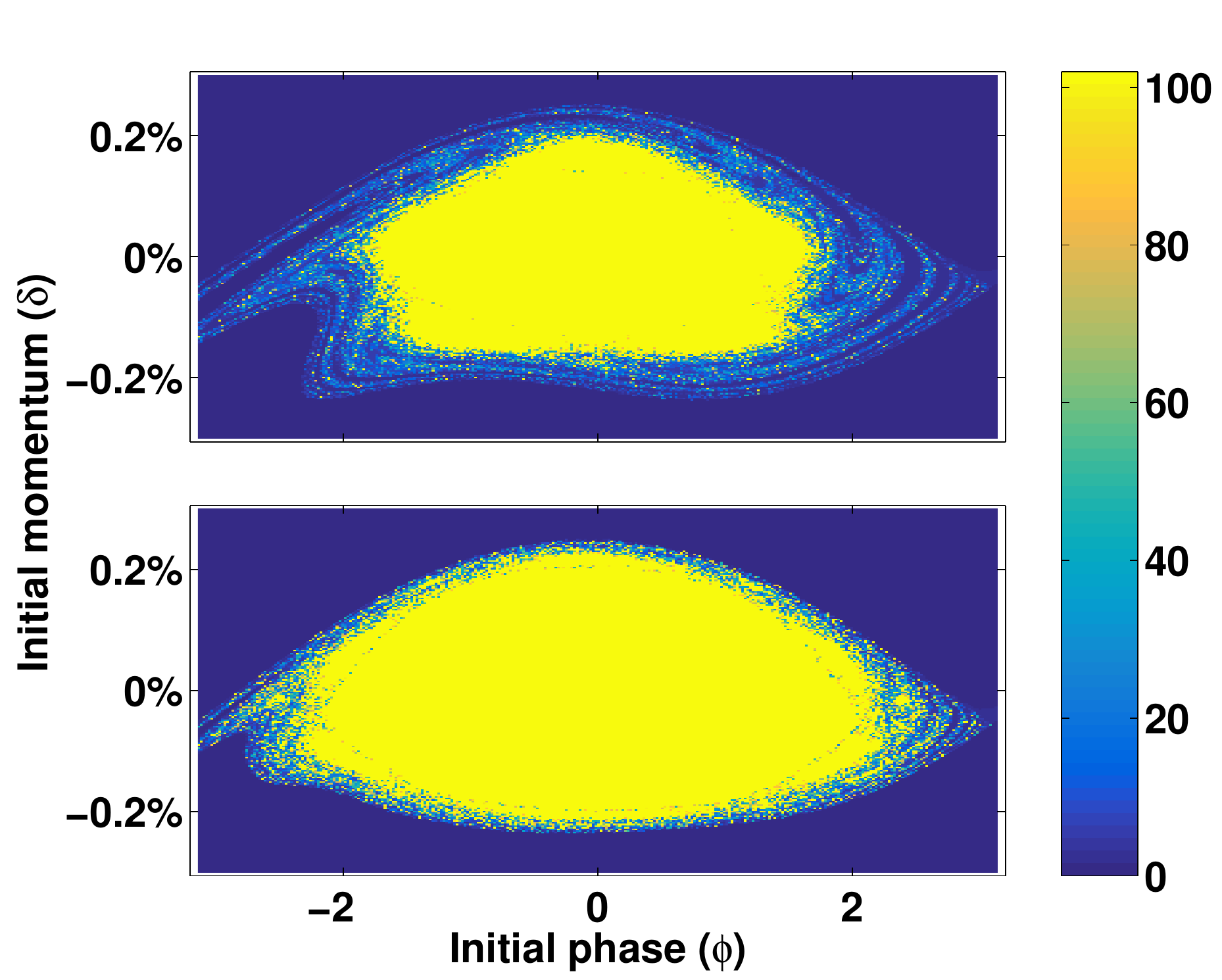}
  \caption{Stability of initial coordinates for $\alpha_{s} = 4.18$. The color shows the number of synchrotron periods a test particle survives before it is lost. The top plot shows conventional slip-stacking and the bottom plot shows harmonic slip-stacking.}
  \label{m}
\end{figure}

We numerically create a stability map \cite{Eldred} for each value of the slip-stacking parameter $\alpha_{s}$ and the harmonic-main voltage ratio $\lambda$. We map the stability of initial particle positions by integrating the equations of motion for each initial position. Each position is mapped independently and only the single particle dynamics are considered. A particle is considered lost if its phase with respect to each of the upper rf cavity, the lower rf cavity, and the average of the two rf cavities, is unbounded. A cut-off phase of $3\pi/2$ is sufficient to classify the trajectory of particles as unbounded. Figure~\ref{m} shows an example of a stability map without a harmonic rf cavity and with a harmonic rf cavity. A selection of harmonic slip-stacking stability maps can be found in Appendix~D of \cite{Dis}.

\begin{figure}[htp]
 \centering
 \begin{overpic}[scale=.4]{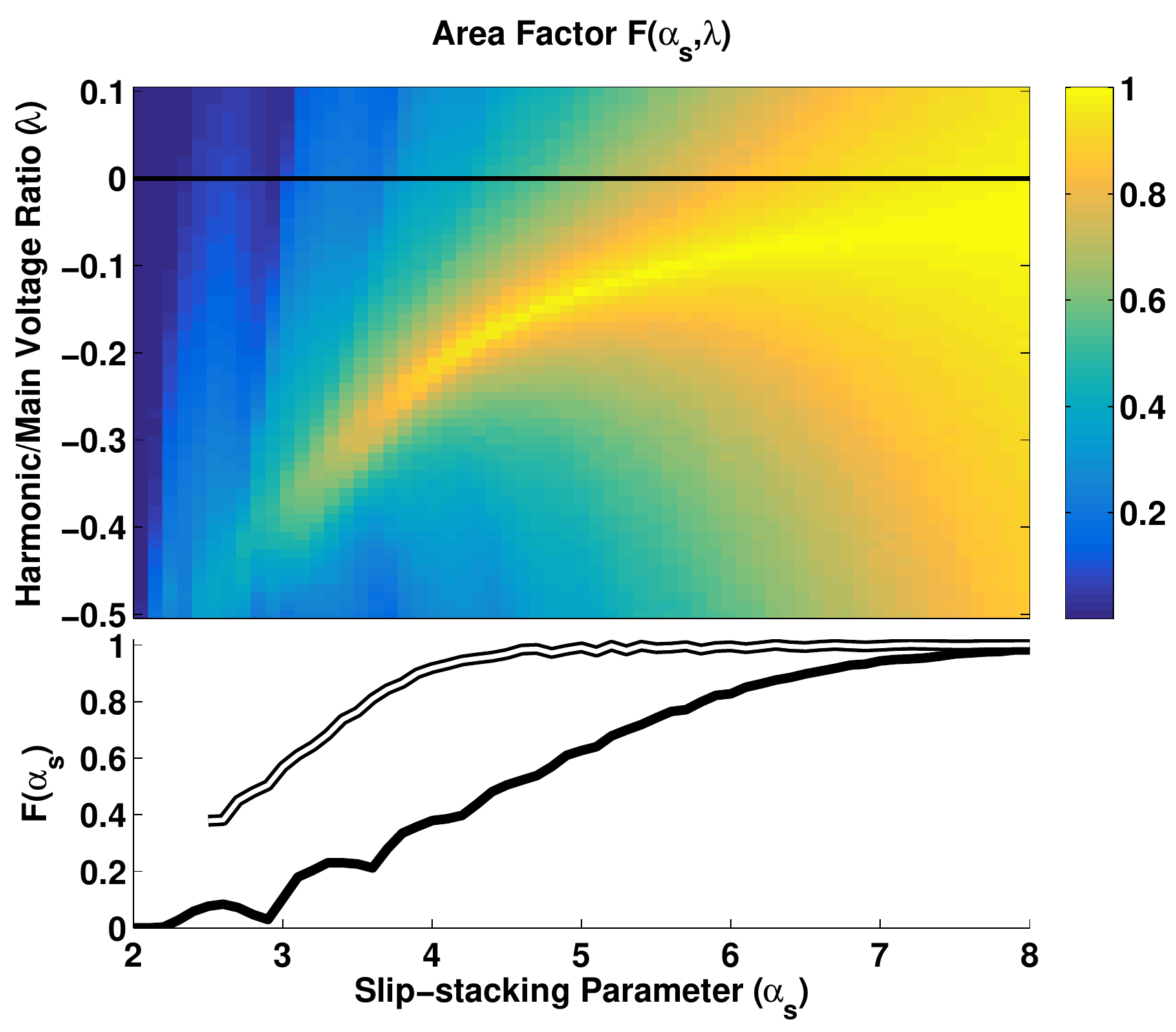}
  \put(47,11){\includegraphics[scale=.3]{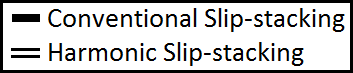}}
 \end{overpic}
 \caption{(top) Slip-stacking area factor $F$ as a function of $\alpha_{s}$ and $\lambda$. (bottom) Slip-stacking area factor $F$ for case without a harmonic cavity ($\lambda = 0$) shown as a single line and case with a harmonic cavity (optimal $\lambda$) shown as a double line.}
 \label{Fmap}
\end{figure}

The bucket area is the product of the total number of ultimately surviving points and the phase-space area sampled by that point. The slip-stacking area factor $F(\alpha_{s},\lambda)$ is the defined to be the ratio between the slip-stacking bucket area to that of a single-rf bucket with the same rf voltage and frequency
\begin{equation} \label{area1}
\mathcal{A}_{s} = \mathcal{A}_{0} F(\alpha_{s},\lambda) = \frac{16}{h |\eta|} \frac{\omega_{s}}{\omega_{rev}} F(\alpha_{s},\lambda).
\end{equation}
The slip-stacking area factor $F(\alpha_{s},\lambda)$ provides a method for calculating the slip-stacking stable phase-space area without requiring each case to be simulated individually~\cite{Eldred}. Figure~\ref{Fmap} shows the slip-stacking area factor $F$ as a function of $\alpha_{s}$ and $\lambda$, with each datapoint calculated from its own stability map.

In application, slip-stacking is tuned by varying $\omega_{s}$ through the main rf voltage while leaving $\omega_{\phi}$ unchanged. The value of $\omega_{\phi}$ is generally constrained by gross features of the accelerators, the Booster harmonic number and ramp rate. We absorb the dependence on $\omega_{s}$ in Eq.~\ref{area1} by defining the modified slip-stacking area factor $Z(\alpha_{s},\lambda)$
\begin{equation} \label{area2}
\mathcal{A}_{s} = \frac{16}{h |\eta|} \frac{\omega_{\phi}}{\omega_{rev}} \left( \frac{F(\alpha_{s},\lambda)}{\alpha_{s}} \right) = \frac{16}{h |\eta|} \frac{\omega_{\phi}}{\omega_{rev}} Z(\alpha_{s},\lambda).
\end{equation}

\begin{figure}[htp]
 \centering
 \begin{overpic}[scale=.4]{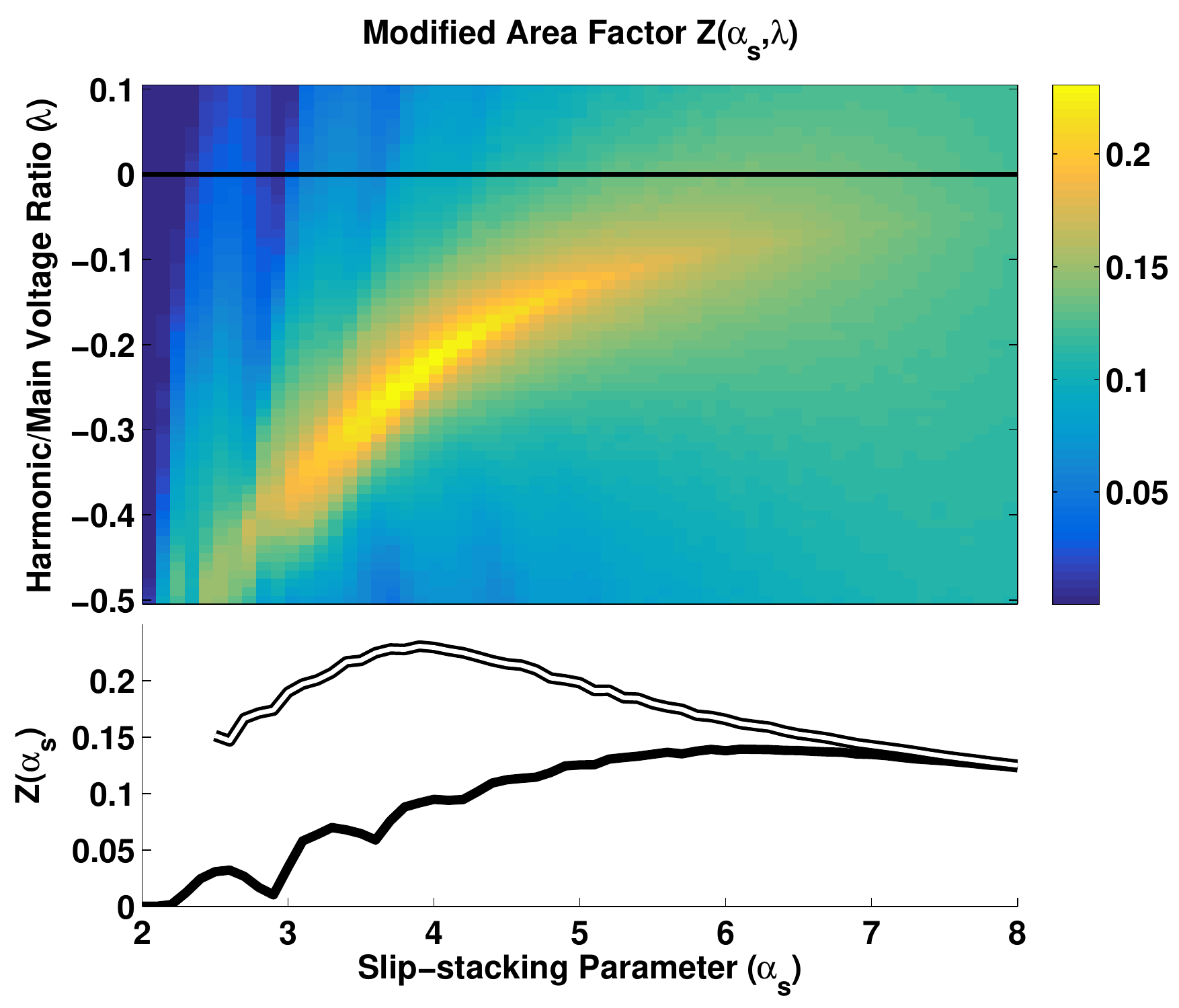}
  \put(47,11){\includegraphics[scale=.3]{EffLegend5.png}}
 \end{overpic}
 \caption{(top) Modified slip-stacking area factor $Z$ as a function of $\alpha_{s}$ and $\lambda$. (bottom) Modified slip-stacking area factor $Z$ for case without a harmonic cavity ($\lambda = 0$) shown as a single line and case with a harmonic cavity (optimal $\lambda$) shown as a double line.}
 \label{Zmap}
\end{figure}

This modified slip-stacking area factor $Z(\alpha_{s},\lambda)$ is proportional to the slip-stacking phase-space area with a coefficient independent of voltage~\cite{Eldred}. Figure~\ref{Zmap} shows the modified slip-stacking area factor $Z$ as a function of $\alpha_{s}$ and $\lambda$.

From Fig.~\ref{Fmap} and Fig.~\ref{Zmap} it is clear that for any value of $\alpha_{s}$, there is an optimal value of $\lambda$ which maximizes the phase-space area. We term this the ``balanced'' condition for $\lambda$. Figure~\ref{Fmap} indicates that for $\alpha_{s} > 4$ at least 90\% of the stable phase-space area can be recovered by using the balanced condition. Figure~\ref{Zmap} indicates that the maximum stable phase-space area with harmonic rf is 65\% higher than that without harmonic rf. $Z(\alpha_{s})$ is maximized at $\alpha_{S} = 3.9$ with harmonic rf and is maximized at $\alpha_{s} = 6.2$ without harmonic rf.

\begin{figure}[htp]
	\centering
    \includegraphics[scale=0.4]{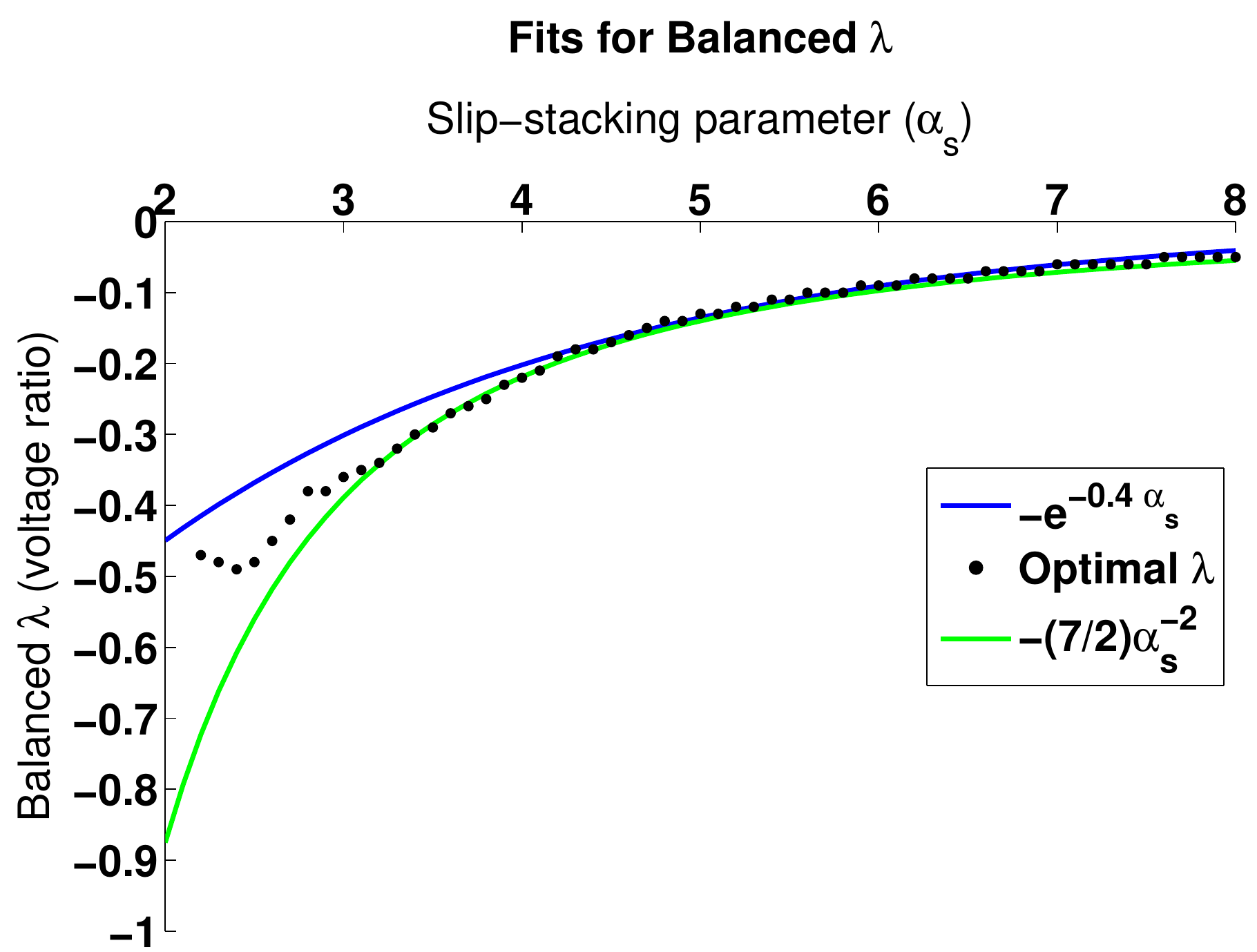}
  \caption{Value of $\lambda$ which maximizes phase-space area, as a function of $\alpha_{s}$.}
  \label{eFit}
\end{figure}

For each value of $\alpha_{s}$, the value of $\lambda$ which maximizes phase-space area is plotted in Fig.~\ref{eFit}. In Fig.~\ref{eFit}, the optimal value of $\lambda$ is fit for $\alpha_{s} > 4$ with Eq.~\ref{fitform}:
\begin{equation} \label{fitform2}
\lambda \approx - e^{-0.4 \alpha_{s}},~\alpha_{s} > 4.
\end{equation}
This fit is consistent with the resonance-canceling mechanism described in the previous section. Figure~\ref{eFit} also shows an empirically-driven equation given by
\begin{equation} \label{engEq}
\lambda \approx - \frac{7}{2} \alpha_{s}^{-2},~\alpha_{s} > 3.
\end{equation}
Equations \ref{fitform2} and \ref{engEq} facilitate application of harmonic slip-stacking by fixing the harmonic rf voltage parameter to a function of the main rf voltage.

\begin{figure}[htp]
 \centering
 \begin{overpic}[scale=.4]{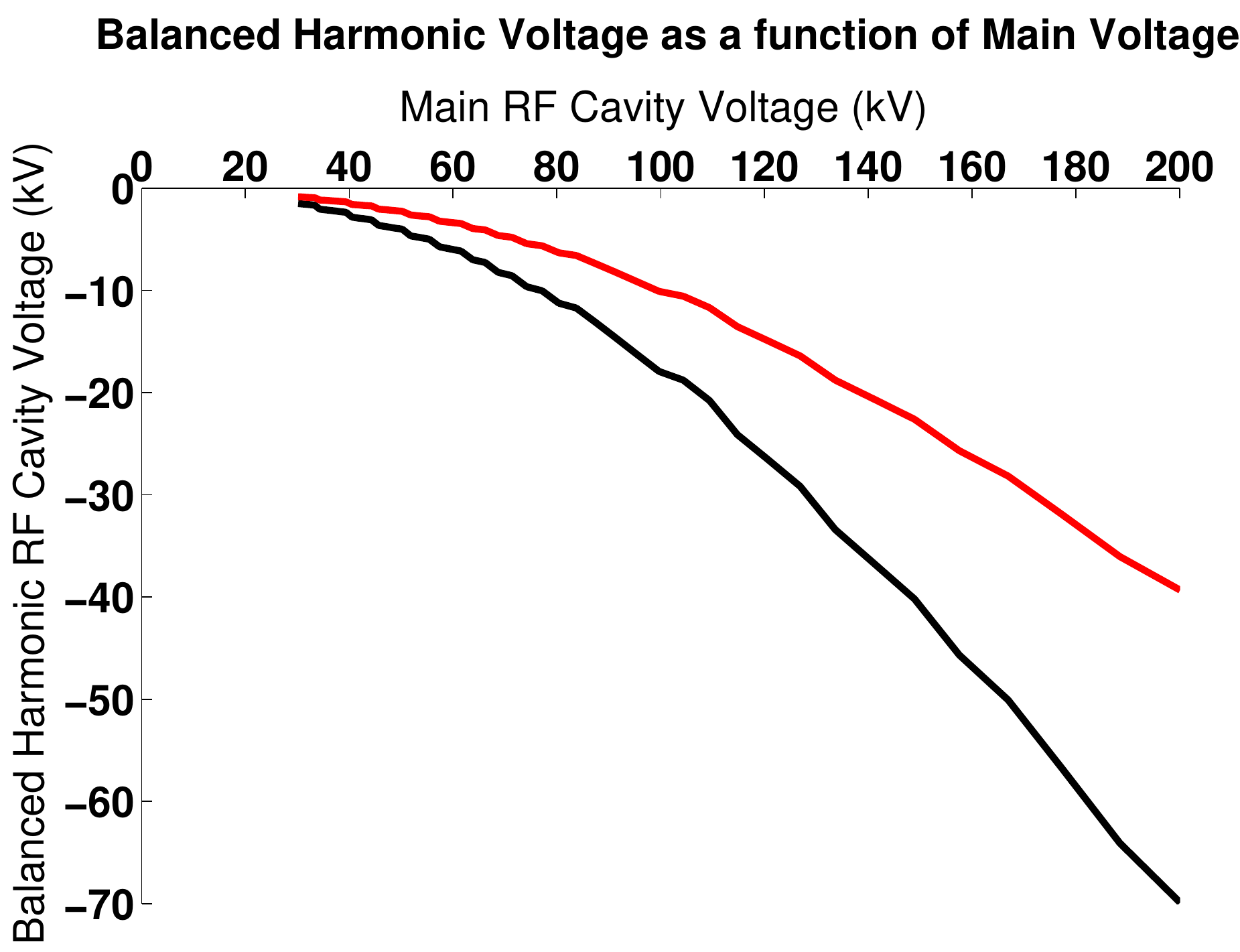}
  \put(18,24){\includegraphics[scale=.3]{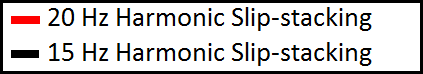}}
 \end{overpic}
 \caption{Balanced value of harmonic rf voltage has a quadratic dependence on main rf voltage. Bottom line shows case for a 15-Hz Booster (black) and top line for 20-Hz Booster (red).}
 \label{empV}
\end{figure}

Recall that the parameter $\lambda$ is defined to be the ratio between the harmonic rf voltage and the main rf voltage $\lambda = V_{H} / V_{M}$. Also, for a fixed rf frequency difference the slip-stacking parameter $\alpha_{s}$ is proportional to the inverse square root of the voltage $\alpha_{s} \propto \omega_{s}^{~-1} \propto V_{M}^{~-1/2} $. If $\lambda$ follows Eq.~\ref{engEq} then the harmonic rf voltage is proportional to the square of the main rf voltage $V_{H} \propto V_{M}^{~2}$. For Recycler parameters (see Table~\ref{Param}) and a 15-Hz Booster cycle-rate, we have
\begin{equation}
{V_{H} \approx - \left(\frac{1.8} {[MV]} \right) V_{M}^{~2}}
\end{equation}
where [MV] represents the megavolt unit. For a 20-Hz Booster cycle rate, we have
\begin{equation}{V_{H} \approx - \left(\frac{1.0} {[MV]} \right) V_{M}^{~2}}.
\end{equation}
Figure~\ref{empV} shows the harmonic rf voltage that maximizes the phase-space area at each main rf voltage.

\begin{table}
\begin{tabular}{| l | l |}
\hline
Recycler Kinetic Energy ($E$) & 8 GeV \\
Recycler Reference rf freq. ($f_{0}$) & 52.8 MHz \\
Recycler Harmonic number ($h$) & 588 \\
Recycler Phase-slip factor ($\eta$) & -8.6*$10^{-3}$ \\
Nom. Recycler rf Voltage ($V_{M}$) & 2 $\times$ 100 kV \\
Booster harmonic number ($h_{B}$) & 84 \\ 
Booster cycle rate ($f_{B}$) & 15/20 Hz \\
Difference in Recycler rf freq. ($\Delta f$) & 1260/1680 Hz \\
\hline
\end{tabular}
\caption{Recycler and Booster parameters used in analysis}
\label{Param}
\end{table}

\section*{Poincar\'{e} Maps}

Poincar\'{e} maps are a traditional tool used to analyze continuous nonautonomous dynamical systems by clearly indicating fixed points, parametric resonances and regions of chaos (see \cite{JSbook} ). The Poincar\'{e} maps presented in this section are obtained by plotting the $\phi$-$\delta$ phase-space coordinates only at every phase-slipping period ($T_{\phi} = 2\pi/\omega_{\phi}$) in a numerical integration of the particle trajectory. We start from 7500 trajectories with initial coordinates uniformly distributed in phase-space and eliminate unbounded trajectories. The Poincar\'{e} maps shown below correspond to a subset of the remaining trajectories that are selected on the basis of approximately even spacing.

\begin{figure}[htp]
	\centering
    \includegraphics[scale=0.4]{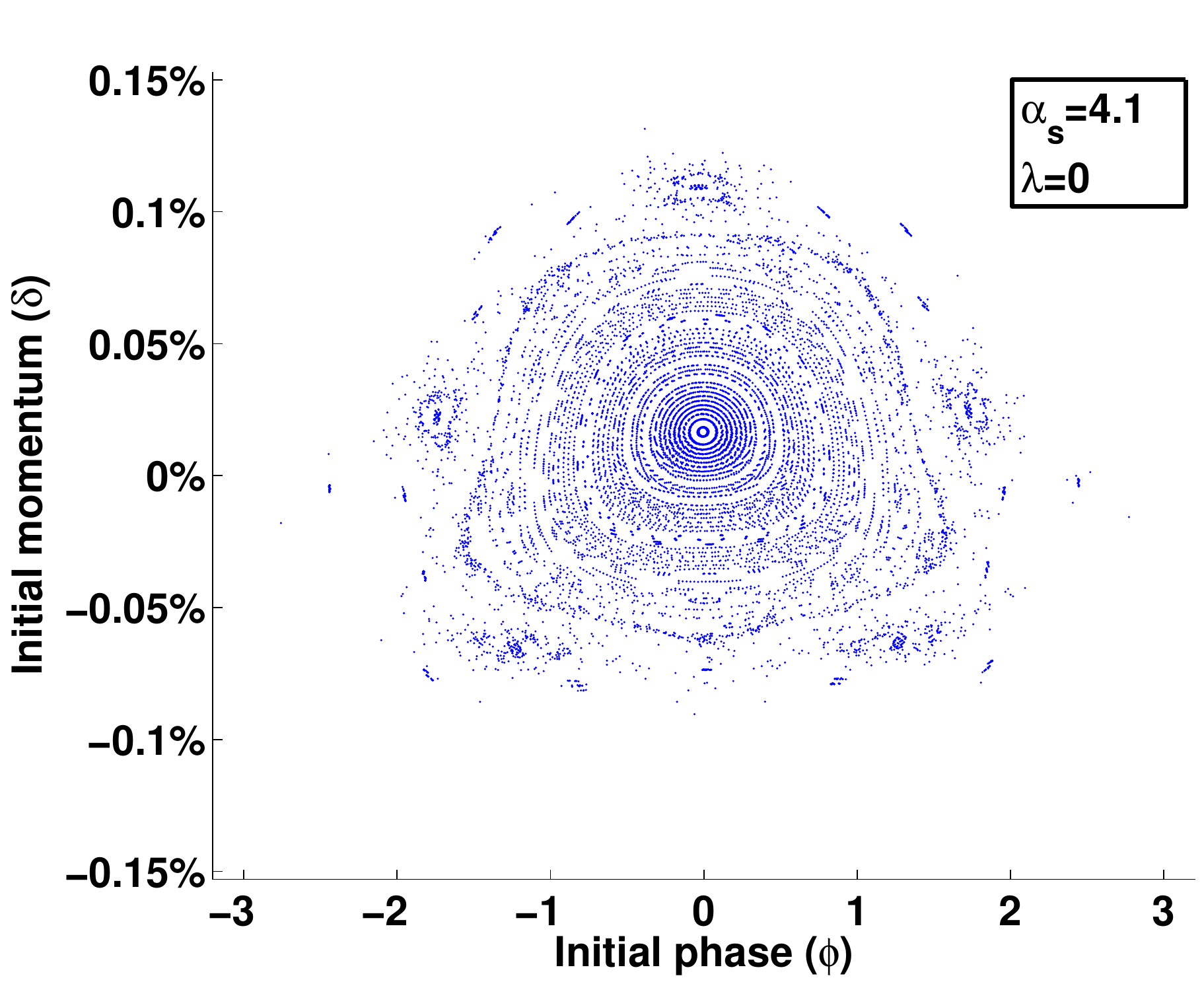}
  \caption{Poincar\'{e} map for conventional slip-stacking with a value of $\alpha_{s}$ corresponding to 100 kV main rf voltage and 1260 Hz rf frequency separation.}
  \label{pncreA}
\end{figure}

\begin{figure}[htp]
	\centering
    \includegraphics[scale=0.4]{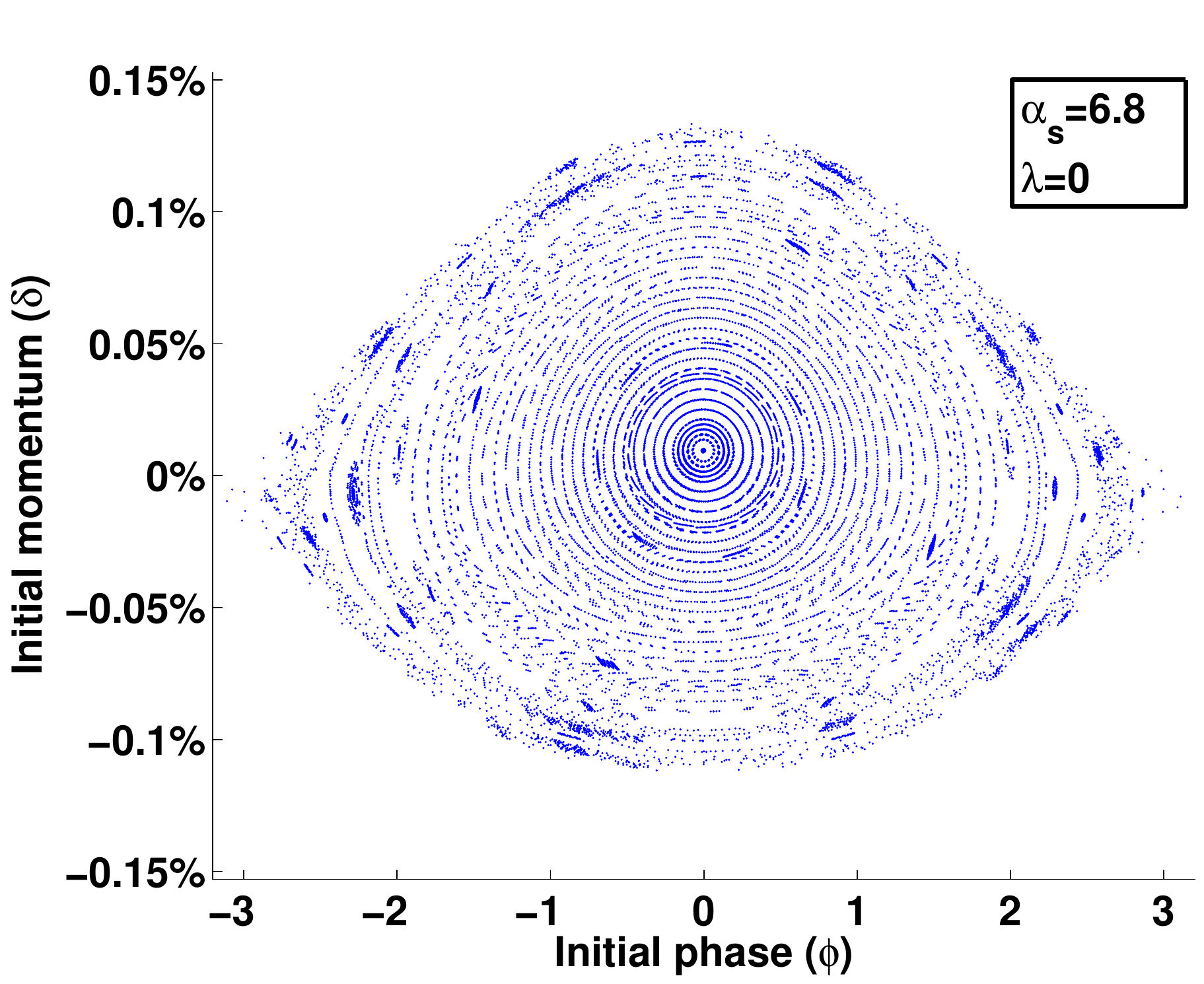}
  \caption{Poincar\'{e} map for conventional slip-stacking with a value of $\alpha_{s}$ corresponding to 65 kV main rf voltage and 1680 Hz rf frequency separation.}
  \label{pncreB}
\end{figure}

\begin{figure}[htp]
	\centering
    \includegraphics[scale=0.4]{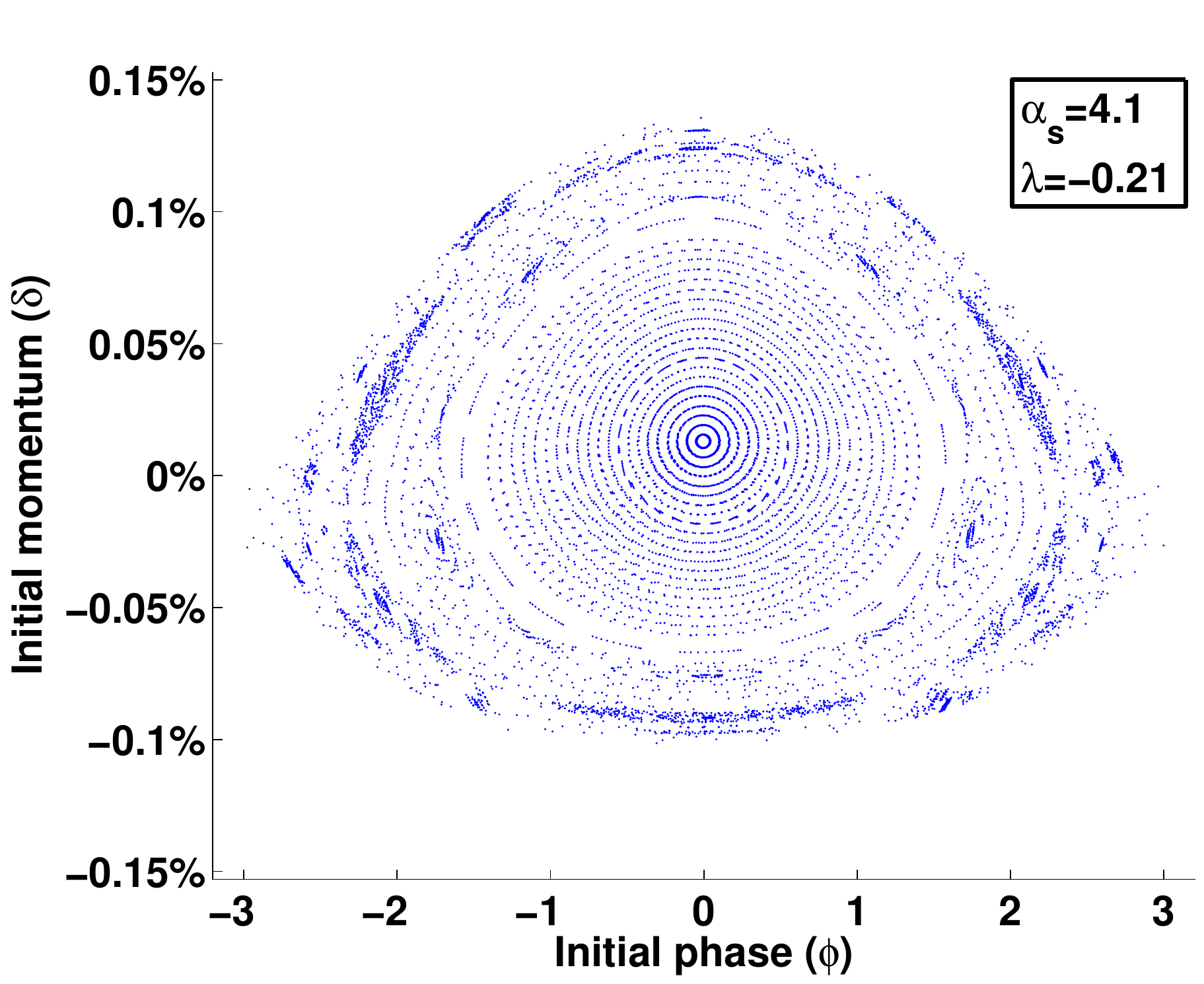}
  \caption{Poincar\'{e} map for harmonic slip-stacking with a value of $\alpha_{s}$ corresponding to 100 kV main rf voltage and a balanced value of $\lambda$ corresponding to 21 kV harmonic rf voltage.}
  \label{pncreC}
\end{figure}

\begin{figure}[htp]
	\centering
    \includegraphics[scale=0.4]{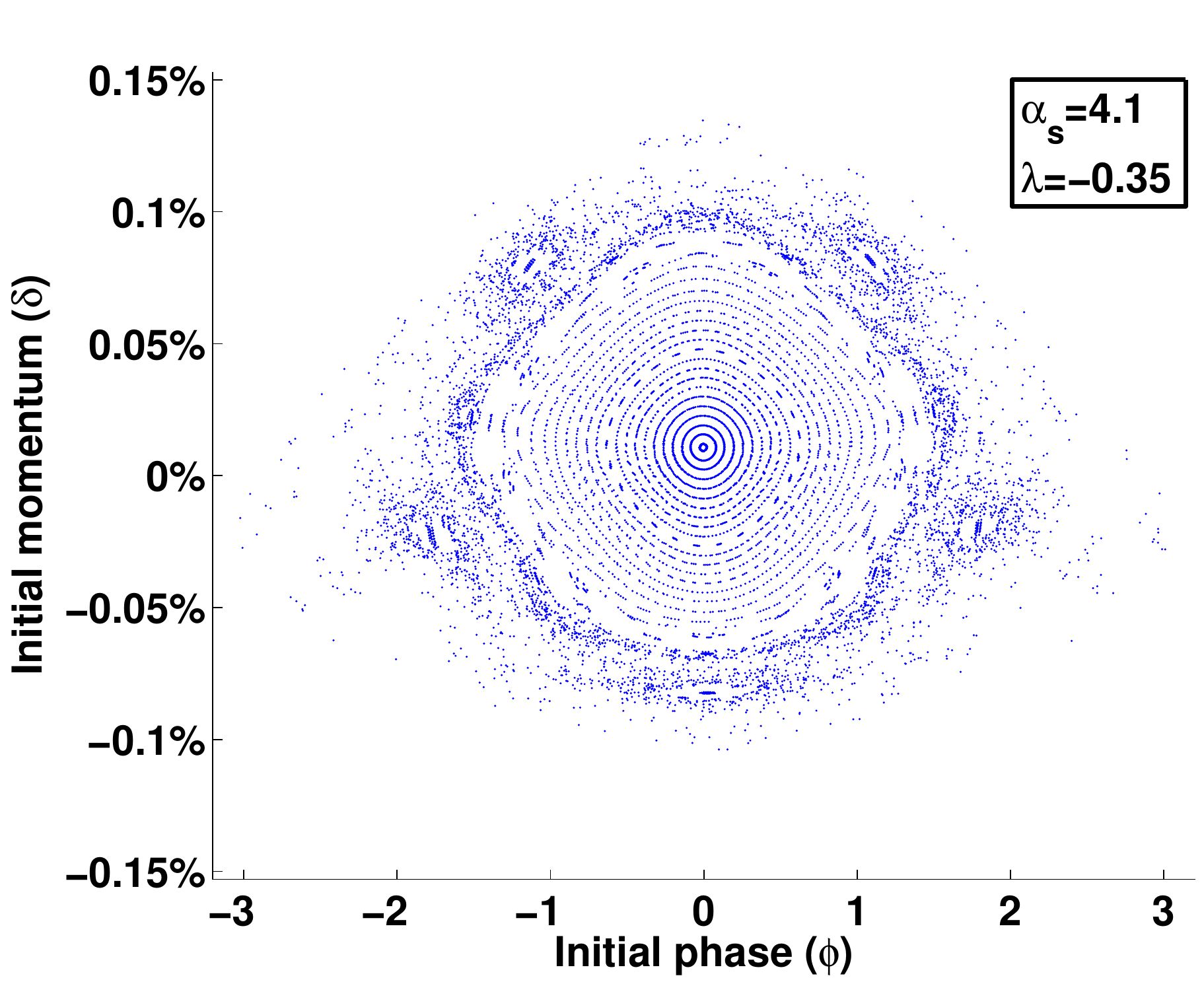}
  \caption{Poincar\'{e} map for harmonic slip-stacking with a value of $\alpha_{s}$ corresponding to 100 kV main rf voltage and an unbalanced value of $\lambda$ corresponding to 35 kV harmonic rf voltage.}
  \label{pncreD}
\end{figure}

Figures \ref{pncreA}--\ref{pncreD} show Poincar\'{e} maps selected to represent the slip-stacking parameter space. Figures \ref{pncreB} and \ref{pncreC} have comparable area factors $F(\alpha_{s},\lambda)$ and both demonstrate large regions of smooth phase-space trajectories. These plots indicate that the harmonic slip-stacking with a balanced value of $\lambda$ reduces impact of the slip-stacking perturbation with a success similar to that of increasing the slip-stacking parameter $\alpha_{s}$. In Fig.~\ref{pncreD} the significant negative value of $\lambda$ has changed the orientation of the fourth-order and fifth-order resonances, as expected from Eq.~\ref{dif2hss}.

A selection of harmonic slip-stacking Poincar\'{e} maps can be found in Appendix~E of \cite{Dis}.

\section*{Injection Efficiency of Gaussian Beams}

The stability maps can also be used to analyze injection scenarios, by weighting the (scaled) stability maps according to a distribution that represents the number of incoming particles injected into that region of phase-space. We used this technique to identify the greatest longitudinal emittance an incoming Gaussian-distributed beam could have and still achieve 99\% injection efficiency at its optimal value of $\alpha_{s}$ and $\lambda$. The 99\% longitudinal beam emittance is given by ${\epsilon_{99\%} = 2.576^{2} \pi \sigma_{p}\sigma_{T}}$.

\begin{figure}[htp]
 \centering
 \begin{overpic}[scale=.4]{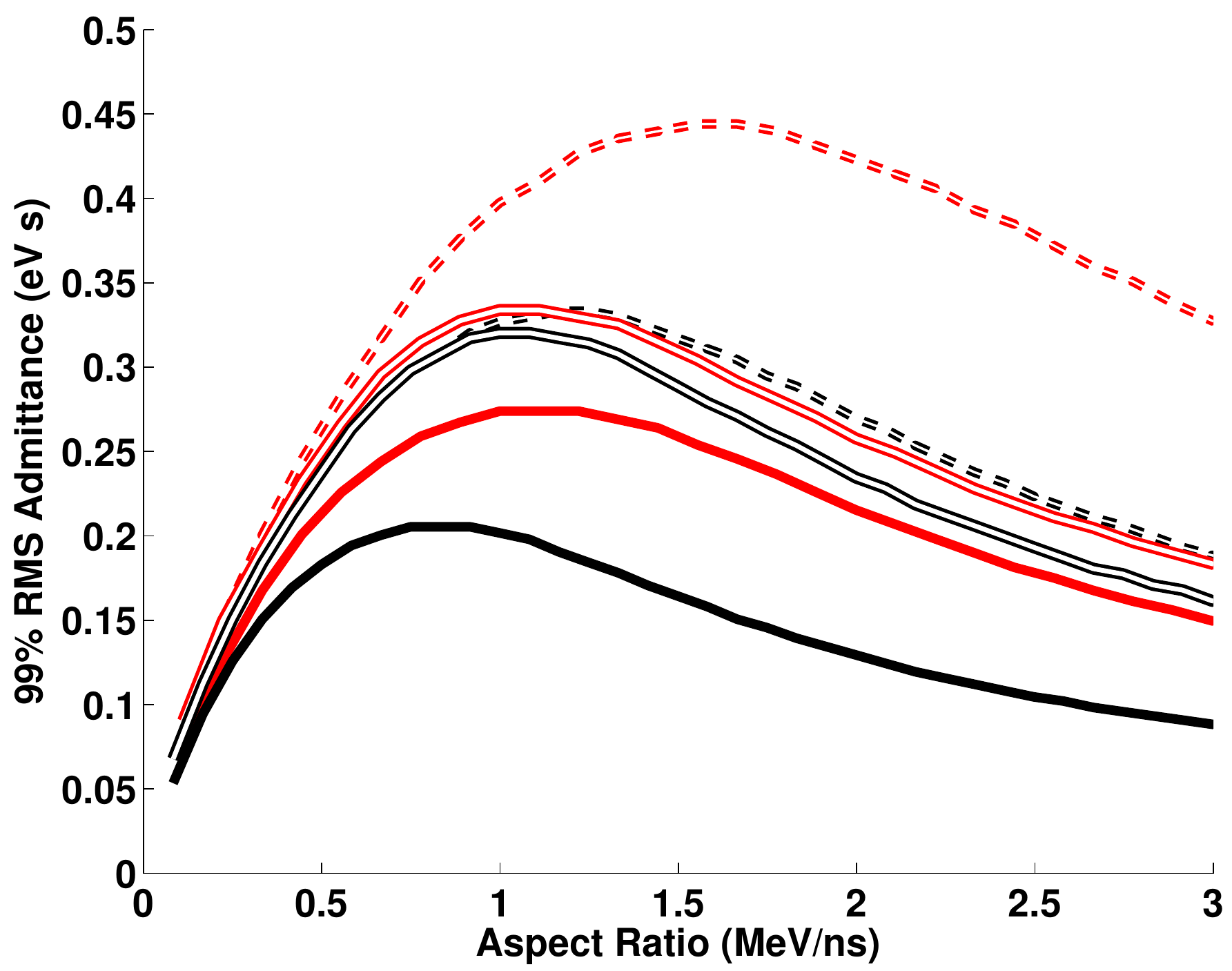}
  \put(20,10){\includegraphics[scale=.3]{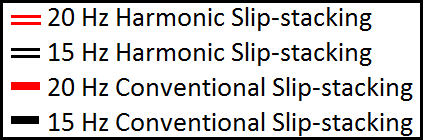}}
 \end{overpic}
 \caption{The 99\% admittance at 99\% efficiency (at an optimal $\alpha_{s}$ and $\lambda$) as a function of aspect ratio. The solid double lines indicate the admittance for harmonic slip-stacking constrained to 100 kV main rf voltage and 20 kV harmonic rf voltage. The dashed double lines indicate admittance for unconstrained rf voltage.}
 \label{Adm}
\end{figure}

Figure~\ref{Adm} shows the 99\% longitudinal admittance as a function of aspect ratio, for conventional slip-stacking, harmonic slip-stacking with constrained rf voltage, and harmonic slip-stacking with unconstrained rf voltage. To achieve the full admittance permitted by 20-Hz harmonic slip-stacking, the main rf voltage must be upgraded to 250 kV and a 70 kV harmonic rf cavity must be installed. However the stable phase-space area provided by this scenario far exceeds the requirements for slip-stacking operation with minimal loss. Instead we propose to keep the main rf voltage constrained to 100 kV and only install a 20 kV harmonic rf cavity. This harmonic slip-stacking scenario is evaluated in greater detail in the next section.

\begin{figure}[htp]
 \centering
 \begin{overpic}[scale=.4]{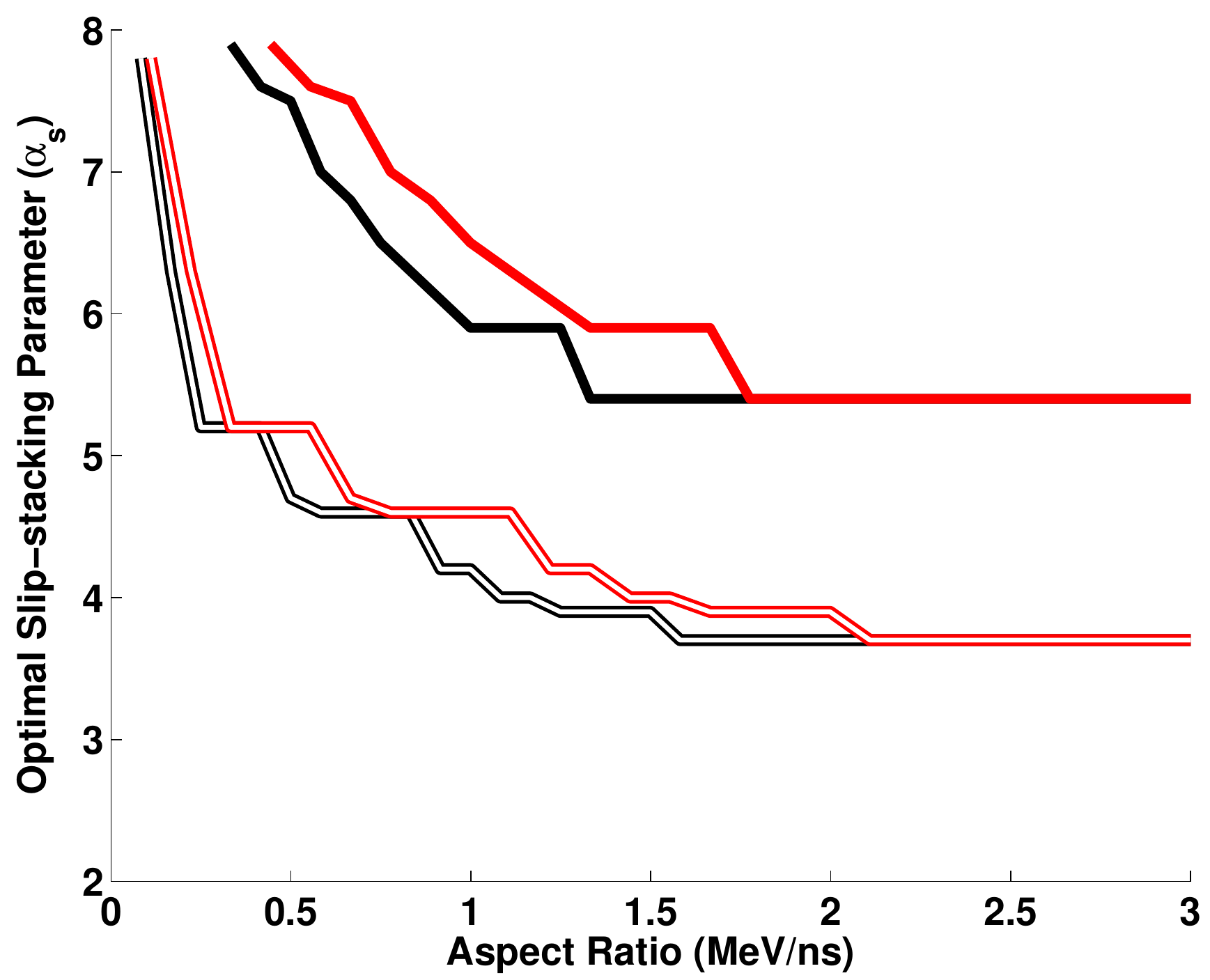}
  \put(53,61){\includegraphics[scale=.3]{EffLegend.png}}
 \end{overpic}
 \caption{Optimal slip-stacking parameter $\alpha_{s}$ for 99\% admittance (at 99\% efficiency) as a function of aspect ratio.}
 \label{BAs}
\end{figure}

\begin{figure}[htp]
 \centering
 \begin{overpic}[scale=.4]{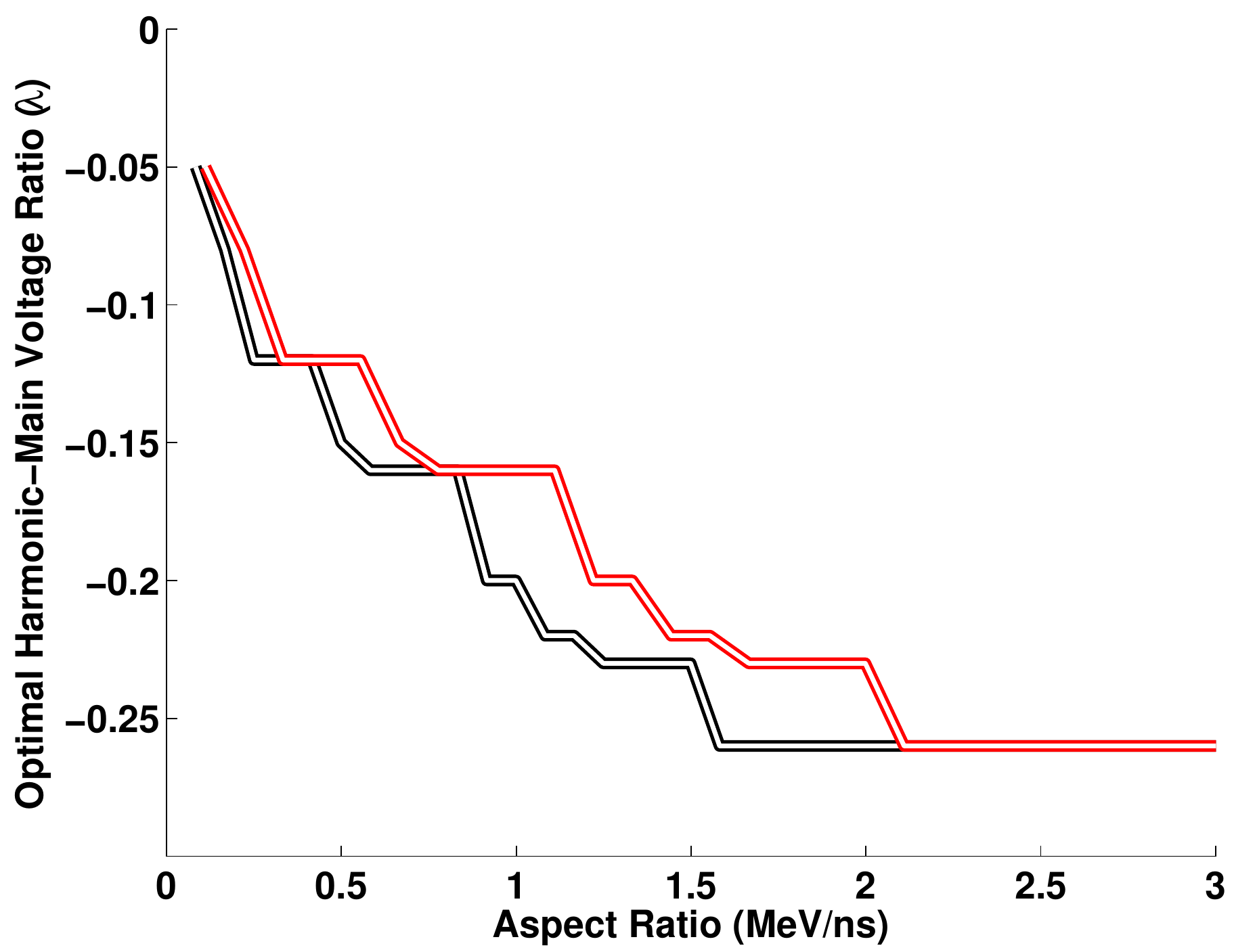}
  \put(53,67){\includegraphics[scale=.3]{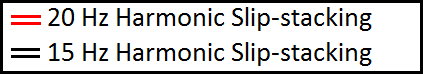}}
 \end{overpic}
 \caption{Optimal harmonic-main voltage ratio $\lambda$ for 99\% admittance (at 99\% efficiency) as a function of aspect ratio.}
 \label{Blam}
\end{figure}

The optimal $\alpha_{s}$ and $\lambda$ at each aspect ratios are shown in Fig.~\ref{BAs} and Fig.~\ref{Blam} respectively (for unconstrained voltage). The value of $\lambda$ with the maximum injection efficiency coincides with the value of $\lambda$ with the balanced condition for maximum stable phase-space area.

The aspect ratio is defined by the momentum spread divided by the temporal spread of the beam distribution. Operationally, the aspect ratio of the beam can be manipulated via bunch rotation in the Booster~\cite{Yang,Ahrens}. A recent measurement found the aspect ratio of the beam injected into the Recycler to be $\sim$1.45 MeV/ns and the 99\% emittance to be 0.25 eV$ \thinspace $s (see next section).

These results were obtaining using parameter values specific to slip-stacking in the Fermilab Recycler (see Table.~\ref{Param}) \cite{Dis}.


\section*{Simulation of Emittance Growth}

After slip-stacking in the Recycler, the two beams are transferred to the Main Injector where they are captured by an rf system operating at a single frequency. Pairs of bunches coalesce into single bunches, with one bunch from each beam captured in the same rf bucket. The longitudinal emittance of the beam after capture in the Main Injector is a critical parameter because it directly impacts losses during transition crossing in the Main Injector. In this section we show that harmonic slip-stacking not also reduces losses directly associated with slip-stacking, but also reduces the emittance of the beam after capture in the Main Injector.

\begin{figure}[htp]
	\centering
    \includegraphics[scale=0.4]{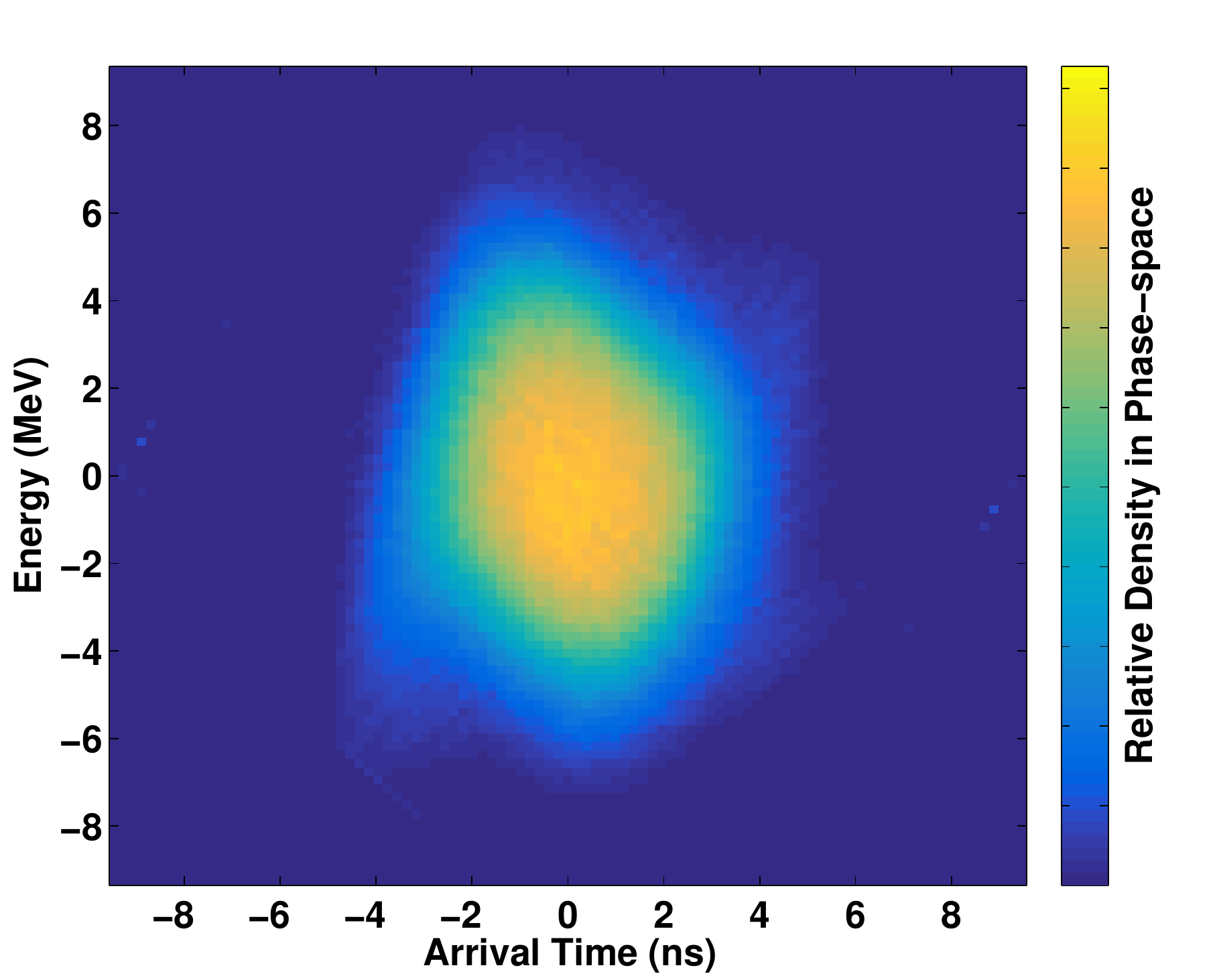}
  \caption{Tomography measurements of longitudinal distribution of beam injected into the Recycler. Distribution derived from an average across Booster batch. Measurement taken from a typical 2+6 slip-stacking cycle on May 27th 2015 at a bunch intensity of $5.1 \times 10^{10}$ protons.}
  \label{Tomo}
\end{figure}

In order to analyze the phase-space distribution of the captured beam, it is important to obtain a realistic model for the phase-space distribution of the injected beam. We used the tomography program developed by Evans~\cite{Evans} to obtain a measurement of the longitudinal distribution of the beam just before slip-stacking in the Recycler. Figure~\ref{Tomo} shows the measurement of the longitudinal distribution of the beam from a typical 2+6 slip-stacking.

The longitudinal distribution obtained from measurement was fit with a bivariate Gaussian distribution. The fit, compared to the direct measurement, has smoother and longer tails. The resulting fit parameters are given in Table~\ref{fParam}. A similar measurement made by Seiya~\textit{et.~al.}~\cite{SeiyaCD} in 2007 found about 70\% greater longitudinal emittance at a comparable beam intensity. 

\begin{table}
\begin{tabular}{| l | l |}
\hline
Temporal Sigma ($\sigma_{t}$) & $\pm$ 2.86 ns \\
Energy Sigma ($\sigma_{E}$) & $\pm$ 4.12 MeV \\
97\% Emittance ($\epsilon_{97\%}$) & 0.17 eV$ \thinspace $s \\
99\% Emittance ($\epsilon_{99\%}$) & 0.25 eV$ \thinspace $s \\
Aspect Ratio ($\sigma_{p}/\sigma_{t}$) & 1.45 MeV/ns \\
\hline
\end{tabular}
\caption{Gaussian fit parameters for longitudinal distribution measured by tomography.}
\label{fParam}
\end{table}

Each bin of the Gaussian distribution was converted to a macroparticle representing the intensity of that bin. The trajectories of these macroparticles were numerically integrated using Eq.~\ref{dif1hss} for 32000 revolutions. Particles with unbounded trajectories were considered lost and removed from the simulation. Next, capture in the Main Injector was simulated with an rf system at a constant voltage and a single frequency. The RMS emittance of the coalesced bunch was calculated by aggregating the particle position over 1000 revolutions. The simulation was repeated under variation of the Recycler main rf voltage, the Recycler harmonic rf voltage, and the Main Injector rf voltage (with the remaining parameters taken from Table~\ref{Param}). For each combination of Recycler voltage parameters, the result with the Main Injector voltage which minimized emittance after capture was selected.

\begin{figure}[htp]
	\centering
    \includegraphics[scale=0.4]{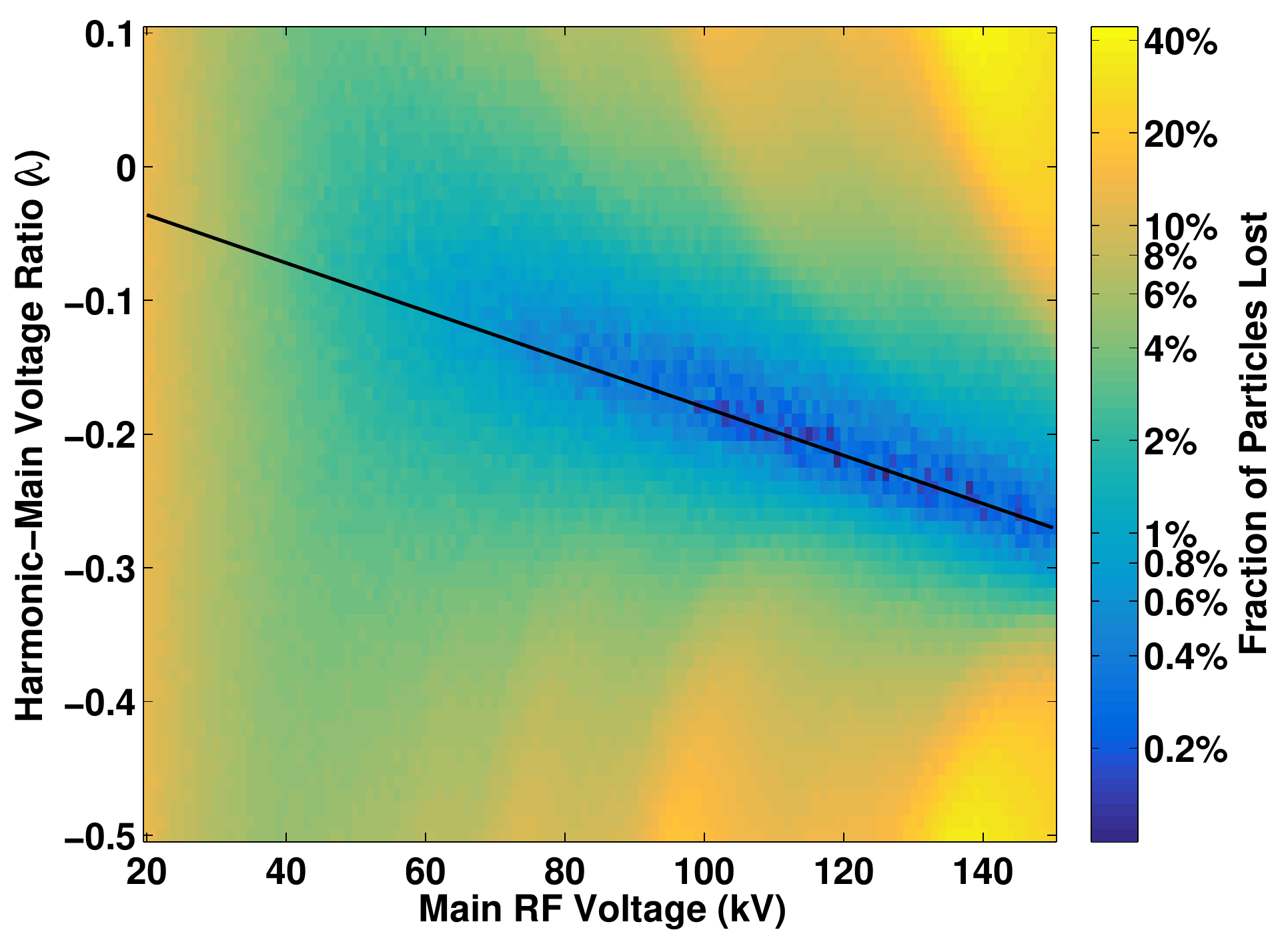}
  \caption{Particle loss as a function of Recycler main rf voltage and $\lambda$.}
  \label{eLoss}
\end{figure}

\begin{figure}[htp]
	\centering
    \includegraphics[scale=0.4]{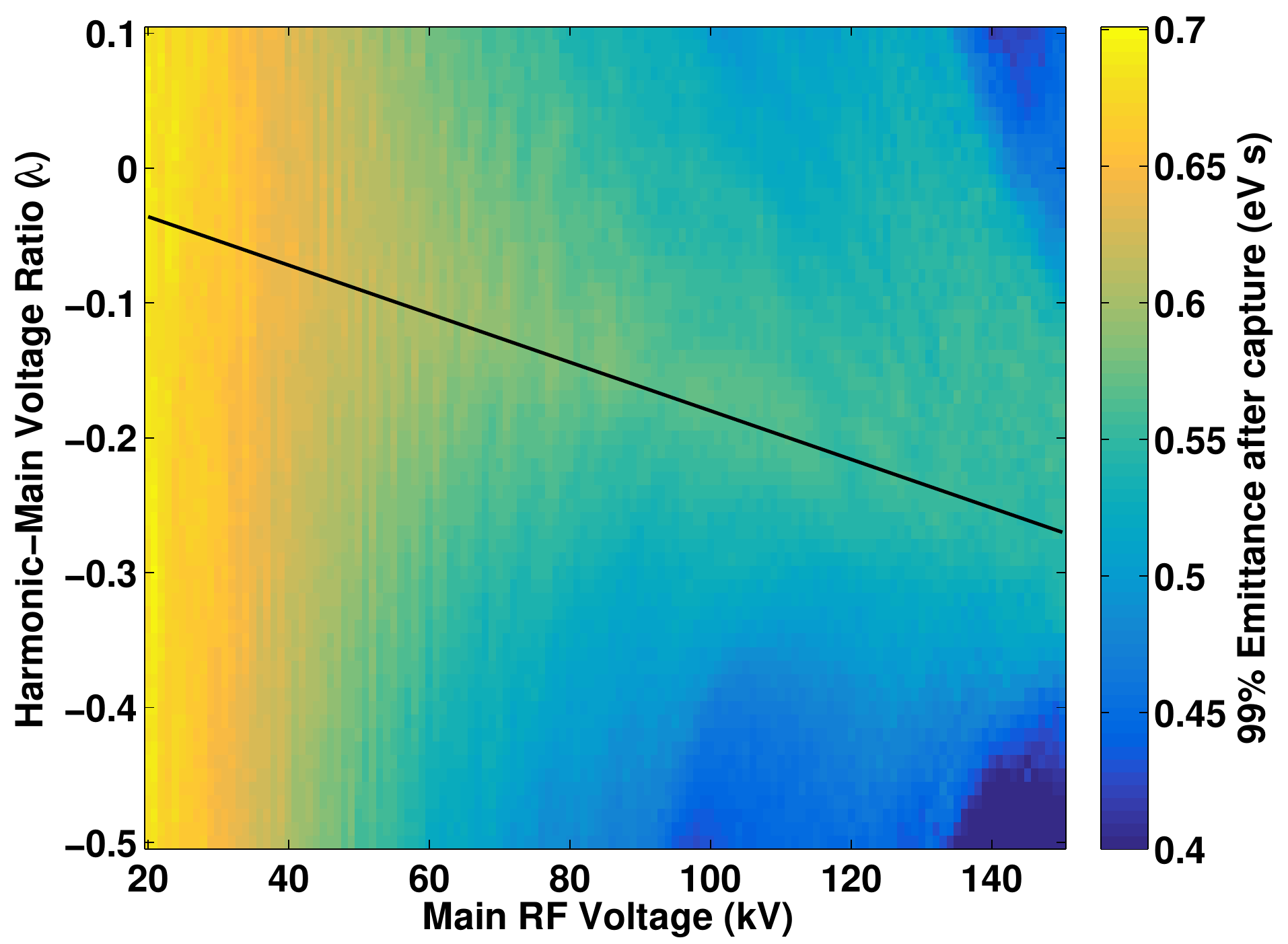}
  \caption{Longitudinal emittance after capture in Main Injector, as a function of Recycler main rf voltage and $\lambda$.}
  \label{eEmit}
\end{figure}

\begin{figure}[htp]
	\centering
    \includegraphics[scale=0.4]{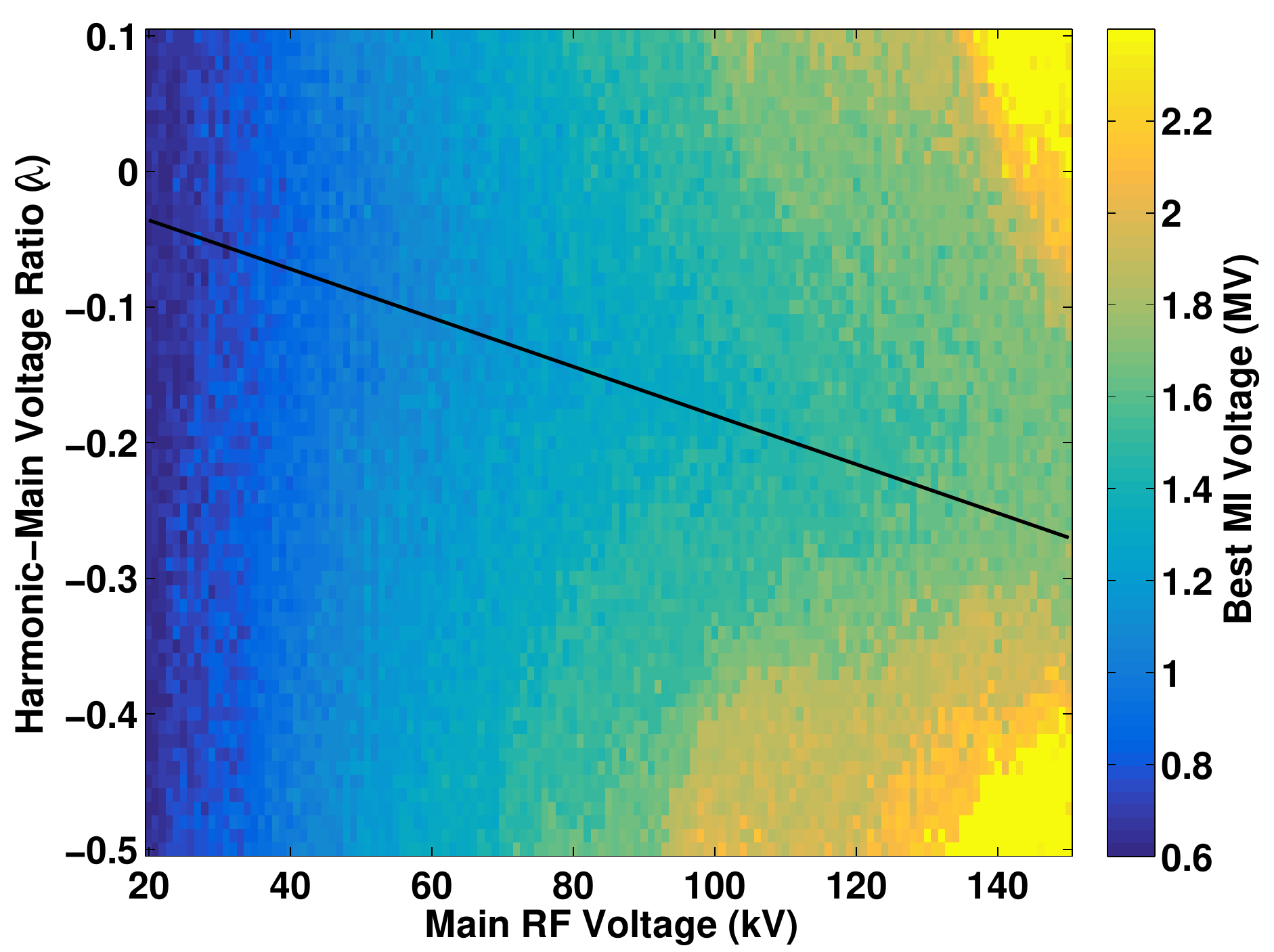}
  \caption{Optimal Main Injector capture voltage as a function of Recycler main rf voltage and $\lambda$.}
  \label{eVolt}
\end{figure}

Figure~\ref{eLoss} shows the particle loss, as a function of the Recycler main rf voltage and harmonic voltage parameter $\lambda$. Particle losses are clearly minimized along the black diagonal line $\lambda \approx -1.8 V_{M}$ where the balanced condition is met. Figure~\ref{eEmit} shows the emittance after capture in the Main Injector, as a function of the Recycler main rf voltage and harmonic voltage parameter $\lambda$. At higher Recycler main rf voltage the aspect ratio is narrower and consequently there is a smaller emittance after capture. Figure~\ref{eVolt} shows the optimal Main Injector rf voltage to minimize emittance.

\begin{figure}[htp]
 \centering
 \begin{overpic}[scale=.4]{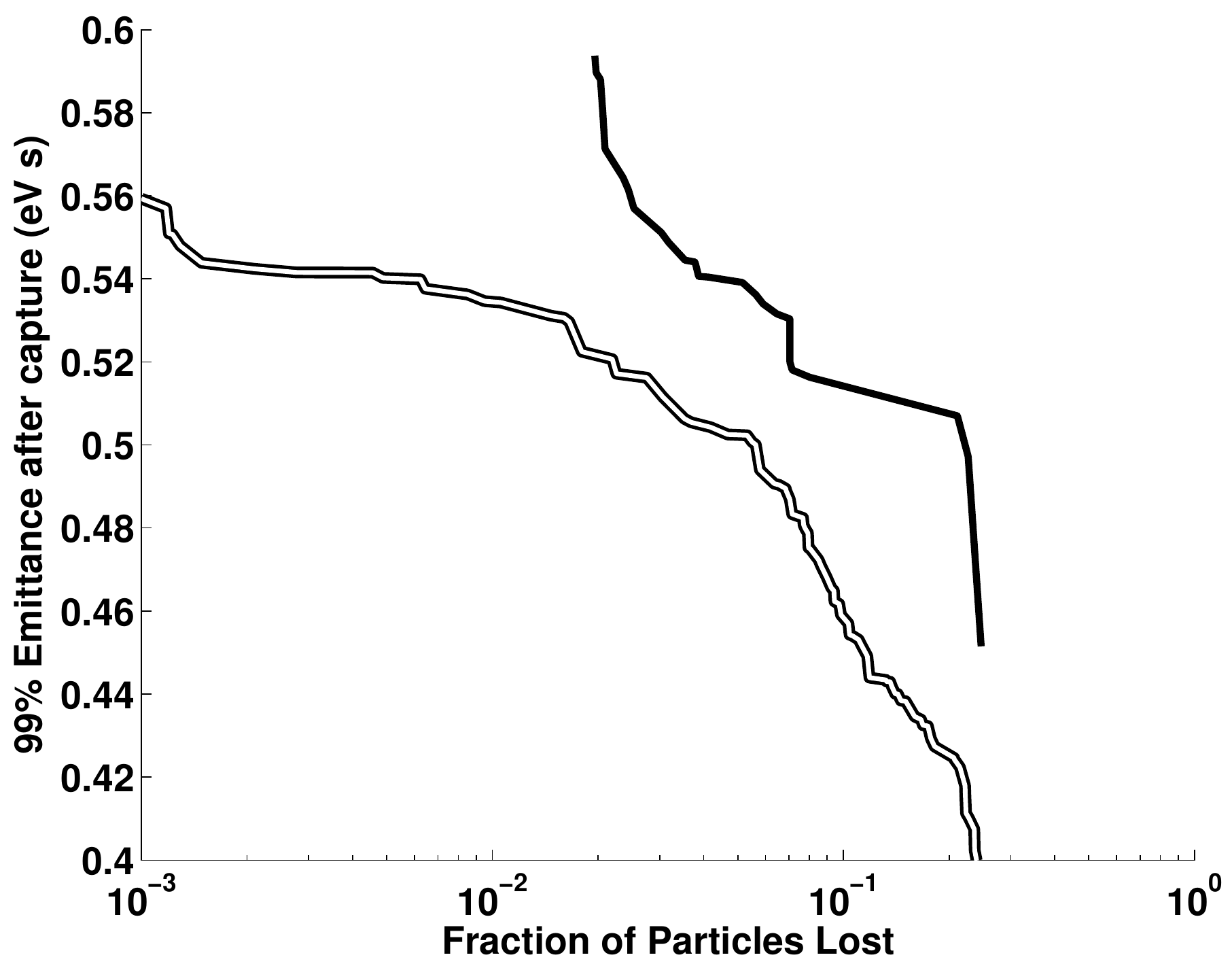}
  \put(53,68){\includegraphics[scale=.3]{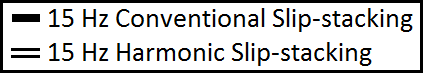}}
 \end{overpic}
 \caption{Pareto fronts for particle loss rate and emittance after capture.}
 \label{Pareto}
\end{figure}

Particle losses during slip-stacking originate from the tail of the longitudinal beam distribution and consequently there is an inherent trade-off between minimizing the loss rate and minimizing the emittance of the beam. The trade-off between these two objectives can be visualized with a Pareto front, the collection of points which minimize one objective while holding the other constant~\cite{DBook}. Figure~\ref{Pareto} shows the Pareto front of particle loss rate and emittance after capture.

By tuning the main rf voltage to minimize particle losses, our simulation indicates conventional slip-stacking can achieve a 2\% loss rate with a corresponding emittance after capture of 0.59 eV$ \thinspace $s. This is consistent with experimentally observed losses during slip-stacking~\cite{Dis}. With harmonic rf, the losses could be reduced to 0.1\% and the corresponding emittance after capture is 0.56 eV$ \thinspace $s. Alternately, losses could be kept at 2\% and the harmonic rf can be used to achieve an emittance after capture of 0.53 eV$ \thinspace $s.

\section*{Conclusion}

We demonstrated how a harmonic cavity can cancel slip-stacking resonances and dramatically increase the dynamical stability of slip-stacking. We derived and verified a model to predict the dependence of the optimal harmonic rf voltage on the main rf voltage. We characterized the stable slip-stacking phase-space area for any combination of accelerator parameters, with and without harmonic rf. We calculated the longitudinal admittance as a function of longitudinal aspect ratio and found the corresponding optimization of parameters $\alpha_{s}$ and $\lambda$. For Gaussian beams, harmonic slip-stacking increases longitudinal admittance 52\% relative to conventional slip-stacking. We measured the longitudinal distribution of the Booster beam with tomography and compared to previous studies. We used the tomography measurement as input for a realistic simulation that calculated the loss-rate, emittance growth, and momentum spread growth of the beam during slip-stacking. From simulation, particle loss can be reduced by a factor of $\sim$20 with a 5\% decrease in longitudinal emittance after capture.

\section*{Acknowledgments}

This work is supported in part by grants from the US Department of Energy under contract DE-FG02-12ER41800 and the National Science Foundation NSF PHY-1205431.


\begin{thebibliography}{99}

  \bibitem{Prebys} E.~J.~Prebys {\sl et al.}, in Proceedings of International Particle Accelerator Conference, 2016, edited by C.~Petit-Jean-Genaz, I.~S.~Ko, K.~R.~Kim, and V.~Schaa.
  
  \bibitem{Adamson} P.~Adamson, in Proceedings of International Particle Accelerator Conference, 2015, edited by S.~Henderson, T.~Satogata, and V.~Schaa.
  
  \bibitem{DUNE} Long-Baseline Neutrino Facility (LBNF) and
Deep Underground Neutrino Experiment (DUNE) Conceptual Design Report, Fermi National Accelerator Laboratory, 2015.
  
  \bibitem{Minos} P.~Adamson {\sl et al.} (MINOS Collaboration), Phys. Rev. Lett. {\bf 110}, 251801 (2013).
  
  \bibitem{Minerva} L.~Fields {\sl et al.} (MINERvA Collaboration), Phys. Rev. Lett. {\bf 111}, 022501 (2013).
    
  G.~A.~Fiorentini {\sl et al.} (MINERvA Collaboration), Phys. Rev. Lett. {\bf 111}, 022502 (2013).
  
  \bibitem{Nova} D.~Ayres {\sl et al.} (NOvA Collaboration), The NOvA Technical Design Report, Fermi National Accelerator Laboratory, 2007.
  
  \bibitem{SeaQuest} P.~E.~Reimer (SeaQuest Collaboration), in Proceedings of International Spin Physics Symposium, 2010, edited by H.~Str\"{o}her and F.~Rathmann.
  
  \bibitem{PIP} P.~Derwent {\sl et al.}, Proton Improvement Plan-II, Fermi National Accelerator Laboratory, 2013.
  
  \bibitem{Dis} J.~Eldred, Ph.D. thesis, Indiana University, 2015.
  
  \bibitem{Eldred} J.~Eldred, R.~Zwaska, Phys. Rev. ST Accel. Beams {\bf 17 }, 094001 (2014).
  
  \bibitem{Derwent} P.~Derwent, S.~Holmes, V.~Lebedev, Fermi National Accelerator Laboratory Report No. BEAMS-DOC-4662, 2014.
  
  \bibitem{BooNE} A.~A.~Aguilar-Arevalo et al. (MiniBooNE Collaboration), Phys. Rev. Lett. {\bf 98}, 231801 (2007).
  
  \bibitem{mu2e} Fermilab Mu2e Conceptual Design Report, 2012.
 
  \bibitem{g-2} B.~L.~Roberts (New Muon (g-2) Collaboration), Chinese Physics C {\bf 34}, 741 (2010). 

  \bibitem{LIU} LHC Injectors Upgrade Technical Design Report Vol. II: Ions, Conseil Europ\'{e}en pour la Recherche Nucl\'{e}aire, 2016
  
  \bibitem{Argyropoulos} T.~Argyropoulos, T.~Bohl and E.~Shaposhnikova, LIU Day Talk, April 11 2014. \url{https://indico.cern.ch/event/299470/session/3/contribution/16/attachments/564153/777111/LIUSlipstacking_2014.pptx}
  
  \bibitem{Valuch} D.~Valuch, in Proceedings of International Conference Radioelektronika, 2009, edited by  O.~Ondr\'{a}\v{c}ek.
  
  \bibitem{Mastoridis} T.~Mastoridis, P.~Baudrenghien, C.~Rivetta, J.~D.~Fox LARP Collaboration Meeting, May 8th 2012. \url{https://indico.fnal.gov/getFile.py/access?contribId=28&sessionId=3&resId=0&materialId=slides&confId=5072}
  
  \bibitem{MacLachlan} J.~A.~MacLachlan, Fermi National Accelerator Laboratory Report No. FN-0711, 2001.
  
  \bibitem{Brown} B.~C.~Brown, P.~Adamson, D.~Capista, W.~Chou, I.~Kourbanis, D.~K~Morris, K.~Seiya, G.~H.~Wu, and M.~J.~Yang, Phys. Rev. ST Accel. Beams {\bf 16}, 071001 (2013).
  
  \bibitem{Mills} F.~E.~Mills, Brookhaven National Laboratory Report No. 15936, 1971.
  
  \bibitem{Boussard} D.~Boussard and Y.~Mizumachi, IEEE Trans. Nucl. Sci. {\bf 26}, 3623 (1979).
  
  \bibitem{SeiyaBC} K.~Seiya, T.~Berenc, B.~Chase, W.~Chou, J.~Dey, P.~Joireman, I.~Kourbanis, J.~Reid, and D.~Wildman, in Proceedings of HB2006, Tsukuba, Japan, 2006, edited by Y.~H.~Chin, H.~Yoshikawa, and M.~Ikegami.
  
  \bibitem{SeiyaB} K.~Seiya {\sl et al.}, in Proceedings of Particle Accelerator Conference, 2005, edited by C.~Horak.
  
  \bibitem{DeyKK} J.~Dey, K.~Koba, I.~Kourbanis, and J.~Reid, in Proceedings of Particle Accelerator Conference, 2007, edited by C.~Petit-Jean-Genaz.
    
  \bibitem{Dey} J.~Dey and I.~Kourbanis, in Proceedings of Particle Accelerator Conference, 2005, edited by C.~Horak.
  
  \bibitem{Madrak} R.~Madrak and D.~Wildman, in Proceedings of North American Particle Accelerator Conference, 2013, edited by T.~Satogata, C.~Petit-Jean-Genaz, and V.~Schaa.
  
  \bibitem{Ainsworth} R.~Ainsworth, P.~Adamson, I.~Kourbanis, and E.~Stern, in Proceedings of International Particle Accelerator Conference, 2016, edited by C.~Petit-Jean-Genaz, I.~S.~Ko, K.~R.~Kim, and V.~Schaa.
  
  \bibitem{Balbekov} V.~Balbekov, Fermi National Accelerator Laboratory Report No. TM-2372-AD, 2007.
  
  \bibitem{SYBook} S.~Y.~Lee, {\it Accelerator Physics}, 3rd Ed. (World Scientific, Singapore, 2012).
  
  \bibitem{JSbook} J.~Jose and E.~Saletan, 1st Ed. {\it Classical Dynamics: A Contemporary Approach}, (Cambridge University Press, Cambridge, United Kingdom, 1998).

  \bibitem{Yang} X.~Yang, A.~I.~Drozhdin, and W.~Pellico, in Proceedings of Particle Accelerator Conference, 2007, edited by C.~Petit-Jean-Genaz.
  
  \bibitem{Ahrens} L.~A.~Ahrens {\sl et al.}, in Proceedings of Particle Accelerator Conference, 1999, edited by A.~Luccio and W.~MacKay.
  
  \bibitem{Evans} N.~J.~Evans, Ph.D. thesis, University of Texas at Austin, 2014.
  
  \bibitem{SeiyaCD} K.~Seiya, B.~Chase, J.~Dey, P.~Joireman, I.~Kourbanis, and J.~Reid, in Proceedings of CARE-HHH-APD Workshop BEAM'07, 2007, edited by W.~Scandale and F.~Zimmermann.
  
  \bibitem{DBook} K.~Deb, 1st Ed. {\it Multi-Objective Optimization using Evolutionary Algorithms}, (John Wiley \& Sons, Hoboken, New Jersey, United States, 2001).
  
\end{thebibliography}
\end{document}